\documentclass[twocolumn]{aastex631}

\usepackage{amsmath}
\newcommand{\msun}{{M_{\odot}}}
\newcommand{\mstar}{{M_{\ast}}}
\newcommand{\ser}{S\'ersic }
\usepackage[utf8]{inputenc}
\newcommand{\LTRpar}{
	\leftskip=0pt plus -.5fil%
	\rightskip=0pt plus .5fil%
	\parfillskip=0pt plus .5fil%
}
\DeclareUnicodeCharacter{202A}{\LTRpar}
\newcommand{\RTLpar}{
	\leftskip=0pt plus .5fil%
	\rightskip=0pt plus -.5fil%
	\parfillskip=0pt plus .5fil%
}
\DeclareUnicodeCharacter{202B}{\RTLpar}
\received{July 18, 2024}
\revised{August 12, 2024}
\accepted{September 26, 2024}

\submitjournal{ApJ}

\shorttitle{‫‪Tracing the Evolution of Intermediate to High-Mass Star-Forming Galaxies}
\shortauthors{Hasheminia et al.}

\begin{document}
	
	\title{‫‪Self-Similar Buildup and Inside-Out Growth: Tracing the Evolution of Intermediate to High-Mass Star-Forming Galaxies Since $z=2$}
	\correspondingauthor{Moein Mosleh}
	\email{moein.mosleh@shirazu.ac.ir}
	
	\author[0000-0003-3428-6441]{Maryam Hasheminia}
	\affiliation{Department of Physics, Institute for Advanced Studies in Basic Sciences (IASBS), 444 Prof. Yousef Sobouti Blvd., Zanjan 45137-66731, Iran}
	\affiliation{Biruni Observatory, College of Science, Shiraz University, Shiraz 71946-84795, Iran}
	\affiliation{Department of Physics, College of Science, Shiraz University, Shiraz 71946-84795, Iran}
	
	\author[0000-0002-4111-2266]{Moein Mosleh}
	\affiliation{Biruni Observatory, College of Science, Shiraz University, Shiraz 71946-84795, Iran}
	\affiliation{Department of Physics, College of Science, Shiraz University, Shiraz 71946-84795, Iran}
	
	\author[0000-0003-3449-2288]{S. Zahra Hosseini-ShahiSavandi}
	\affiliation{Biruni Observatory, College of Science, Shiraz University, Shiraz 71946-84795, Iran}
	\affiliation{Department of Physics and Astronomy, University of Padua, Vicolo Osservatorio 3, 35122 Padova, Italy}
	
	\author[0000-0002-8224-4505]{Sandro Tacchella}
	\affiliation{Kavli Institute for Cosmology, University of Cambridge, Madingley Road, Cambridge, CB3 0HA, UK}
	\affiliation{Cavendish Laboratory, University of Cambridge, 19 JJ Thomson Avenue, Cambridge, CB3 0HE, UK}
	
	
	\begin{abstract}
		We aim to discern scenarios of structural evolution of intermediate to high-mass star-forming galaxies (SFGs) since cosmic noon by comparing their stellar mass profiles with present-day stellar masses of $\log(M_{\ast,0}/\msun)=10.3-11$. We addressed discrepancies in the size evolution rates of SFGs, which may be caused by variations in sample selection and methods for size measurements. To check these factors, we traced the evolution of individual galaxies by identifying their progenitors using stellar mass growth histories (SMGHs), integrating along the star-forming main sequence and from the IllustrisTNG simulations. Comparison between the structural parameters estimated from the mass- and light-based profiles shows that mass-weighted size evolves at a slower pace compared to light-based ones, highlighting the need to consider the mass-to-light ratio ($M/L$) gradients. Additionally, we observed mass-dependent growth in stellar mass profiles: massive galaxies ($\log(M_{\ast,0}/\msun)\gtrsim10.8$) formed central regions at $z\gtrsim1.5$ and grew faster in outer regions, suggesting inside-out growth, while intermediate and less massive SFGs followed a relatively self-similar mass buildup since $z\sim2$. Moreover, slopes of observed size evolution conflict with the predictions of TNG50 for samples selected using the same SMGHs across our redshift range. To explore the origin of this deviation, we examined changes in angular momentum (AM) retention fraction using the half-mass size evolution and employing a simple disk formation model. Assuming similar dark matter halo parameters, our calculations indicate that the AM inferred from observations halved in the last 10 Gyr while it remained relatively constant in TNG50. This higher AM in simulations may be due to the accretion of high-AM gases into disks. 
	\end{abstract}
	
	\section{Introduction} \label{sec:intro}
	Despite the notable progress achieved in extragalactic astrophysics, questions persist regarding the construction of galaxies' structural components and the mechanisms involved. It is known that star-forming galaxies (SFGs) assemble their stellar mass through star formation (SF) which is regulated by feedback mechanisms, accretion of gas, and minor mergers. Furthermore, various internal and external processes, such as bar and disk instability and clump migration can transfer or redistribute the stellar mass \citep{Noguchi1999, Kauffmann2004, Kormendy2004, Bournaud2007, Genzel2008, Dekel2009, Krumholz2010}. Nevertheless, how these processes operate on distinct regions of galaxies and which ones are more effective across different masses and redshifts is still debated.
	
	Each internal and external mechanism leaves distinctive imprints on the stellar mass distribution within galaxies and consequently, by investigating the assembly history of the stellar mass within galaxies, it becomes possible to impose better constraints on existing galaxy formation and evolution scenarios. Therefore, many studies have explored the stellar mass buildup in SFGs over cosmic time and showed that in general, SFGs and particularly, the high-mass ones accumulate their stellar mass from the central regions outward, namely the inside-out growth \citep{vanDokkum2010, Szomoru2011, Patel2013a, Hilz2013, vanDokkum2015, Nelson2016, Hill2017, Abdurrouf2023}. The main evidence provided for this model is the notably smaller size of high redshift galaxies compared to their local counterparts, as well as centrally concentrated stellar mass surface density profiles of galaxies directly derived from their light distribution. In contrast, some observations have indicated a slower evolution in the size of SFGs compared to quiescent galaxies (QGs) at a fixed mass and mainly at redshifts $\lesssim 1.5-2$ \citep{Barden2005, Trujillo2006, Curtis-Lake2016, Mosleh2017, Jimenez2019, Suess2019a, Mosleh2020, Ji2023}. In some studies, the little evolution in size is argued to be related to the moderately flat specific star formation rate (sSFR) profiles obtained from observations and simulations, hence showing that the assembly of their stellar contents is relatively self-similar \citep{Patel2013b, vanDokkum2013, Liu2017, Wang2017, Tacchella2018, Suess2019b, Nelson2019, Nelson2021}. The ``self-similar growth" scenario suggests mass builds up simultaneously across all radii at approximately the same rate through mainly star formation activities. These diverse scenarios emphasize the need for more detailed studies on the stellar mass growth history (SMGH) of SFGs to discern the dominant growth pattern in each mass regime and the underlying mechanisms that drive it.
	
	The discrepancy in the interpretation of different scenarios is mainly based on the variation in measurements of the structural parameters such as size and the \ser index. This can be exemplified in a wide range of slopes that has been measured in literature for the size evolution of SFGs as a function of redshift i.e., $r\propto(1+z)^\gamma$, in which $\gamma$ spans from $\sim0$ to -3. Various studies reported a significant growth in the size of SFGs, with slopes ranging from -1 \citep{Bouwens2004, Dahlen2007, Williams2010, Huang2013, Holwerda2015, Papovich2015, Yang2021} to -1.5 \citep{Ferguson2004, Hathi2008}, or somewhere in between \citep{Oesch2010, vanDokkum2010, Mosleh2011, Mosleh2012, Ono2013, Shibuya2015, Kawamata2018, Lindroos2018, Shibuya2019}. Interestingly, even larger rates are predicted in some simulations and semi-analytical models \citep{Liu2017, Marshall2019, Roper2022}. On the other hand, many studies have found that the slope of size evolution can be comparatively shallower, between -0.5 to -0.9 \citep{Buitrago2008, Franx2008, Bruce2014, Morishita2014, vanderWel2014, Allen2017, Paulino-Afonso2017, Mowla2019, Gomez-Guijarro2022, Ormerod2024}, or even lower \citep{Trujillo2006, Mosleh2017, Mosleh2020, Jimenez2019, Hasheminia2022, Ji2023, Morishita2024}. Such discrepancy in the rate of size evolution can be attributed to two primary factors: the approach taken to choose a sample and the way that the sizes are measured.
	
	To accurately comprehend the growth of individual galaxies, it is crucial to reconstruct the evolutionary path of each galaxy, which observationally is challenging. Several approaches have been suggested to establish a connection between progenitors and descendants and trace the potential pathway of evolution for each galaxy. The most widely employed approach is reconstructing the mass growth history using number density arguments \citep{vanDokkum2010, vanDokkum2013, Patel2013a, Whitney2019}, abundance matching \citep{Papovich2015}, and the evolution of the SFR-mass relation \citep{Patel2013b, Hasheminia2022, George2024}. Using the constant number density method, \cite{Patel2013a} determined a slope of 0.63 for SFGs with a final mass of $10^{11.2}\msun$. Similar patterns of slower growth can also be seen in other studies that rely on progenitor selection \citep{Patel2013b, vanDokkum2013, Whitney2019, Hasheminia2022, Ji2023, George2024}, whereas steeper slopes were reported for samples selected at fixed mass \citep{Buitrago2008, vanderWel2014, Allen2017, Faisst2017, Mowla2019}. Over a similar redshift range as \cite{Patel2013a}, \cite{vanderWel2014}, and \cite{Mowla2019} found mass-dependent slopes of 0.8 and 0.9 for the size evolution of SFGs at the fixed mass bins of $\log(\mstar/\msun)=[11-11.5]$ and $[10.75-11.3]$. The progenitor bias, as evidenced in \cite{Ji2023}, is probably responsible for these higher rates.
	
	In addition, the obtained size growth rate is significantly influenced by the observing band, as different wavelengths are more sensitive to distinct stellar populations. UV and shorter optical wavelengths are generally more sensitive to young blue populations, whereas longer optical and NIR wavelengths are suitable for tracing the old red components, leading to variations in structural parameters based on wavelength \citep{Peletier1996, Mollenhoff2006, Graham2008, LaBarbera2010, Vulcani2014, Lange2015, Jimenez2021}. Moreover, dust attenuation and metallicity gradient, often present in early-type galaxies, can affect the galaxy's flux, so its size and shape may be changed in different observing bands \citep{Pierini2004, Tuffs2004, Mollenhoff2006, Graham2008, LaBarbera2010, Kelvin2012, Vulcani2014, Lange2015}. A color gradient is caused by these factors, which result in radial mass-to-light ratios ($M/L$) in galaxies. Due to the radial variations in $M/L$, especially in SFGs, the light distribution strongly depends on the observed wavelength and may not coincide with the mass distribution. Previous studies indicate that galaxies, on average, tend to have a negative color gradient, implying that UV/Opt radii (hereafter half-light) are often larger than mass-based/NIR ones and decrease with increasing wavelengths \citep{Szomoru2013, Wuyts2013, Chan2016, Liu2017, Mosleh2017, Wang2017, Suess2019a, Miller2023, vanderWel2024}. Furthermore, the ratio of $R_{e,mass}/R_{e,light}$ generally declines towards lower redshifts, introducing a substantial discrepancy between half-mass/NIR radii and optical half-light radii at later cosmic times, especially for higher mass galaxies \citep{Mosleh2017, Suess2019a, Suess2019b, Ibarra-Medel2022, Miller2023, vanderWel2024}. 
	
	Recent studies have shown that the redshift variations of $R_{e,mass}/R_{e,light}$ can alter the evolutionary trend, causing mass-based radii to predict slower size growth than light-based radii \citep{Mosleh2017, Suess2019a, Suess2019b, Mosleh2020, Hasheminia2022, Miller2023}. Fortunately, the availability of data from NASA's James Webb Space Telescope (JWST) enables the examination of not only the structure of galaxies but also the robustness of mass-based parameters estimated from the HST images. \cite{vanderWel2024} demonstrated that average half-light sizes are up to 40\% smaller in NIR wavelengths than in optical bands, while NIR radii agree well with half-mass sizes. Furthermore, examining JWST data has revealed that SFGs' sizes evolve at a slower pace in NIR relative to UV/Opt \citep{Morishita2024, Ward2024, Varadaraj2024}. \cite{Varadaraj2024} measured a slope of 0.6 for LBGs at $3<z<5$, which is aligned with the slow growth in sizes of LBGs observed by \cite{Morishita2024} across a wide range of $z=5-14$.
	
	Given the existing disagreements, it becomes crucial to carefully consider the selection method and measurement approach when examining the size evolution of galaxies. Moreover, some recent studies have suggested that the evolution of half-mass sizes appears to be more rapid in simulations compared to that estimated from observations \citep{Suess2019b, vandeSande2019, Hasheminia2022, Miller2023}. Therefore, it is imperative to undertake more extensive analyses to examine this discrepancy between observations and simulations. Thus, we were motivated to explore the size evolution of SFGs by tracking the variations in progenitors' mass-based radii over different redshifts. In the present study, we attempt to explore the stellar mass accumulation and size evolution of SFGs by extending our previous work to the intermediate stellar mass range of $10^{10.3} - 10^{11} \msun$. To achieve this, we utilize the structural parameters, such as semi-major and circularized radii and \ser index, along with the stellar mass surface density and sSFR profiles. These parameters are estimated from the stellar mass maps, allowing for a more direct comparison between observations and simulations. To properly analyze the evolution of each galaxy, we endeavor to trace it over time by identifying probable progenitors from high redshift to the local universe. 
	
	This paper is set out as follows: Section \ref{sec:method} describes the data set and how we trace the progenitors and select the samples. The analysis of the evolution of mass-based structural parameters and profiles of the stellar mass and SFR are presented in Section \ref{sec:result}. Finally, we discuss and summarize our results in Sections of \ref{sec:discussion} and \ref{sec:summary}. Throughout this paper, we employed the standard cosmological parameters of $\Omega_{m}$ = 0.3, $\Omega_{\Lambda}$ = 0.7, and $H_{0}$ = 70 $\mathrm{km\ s}^{-1} \mathrm{Mpc}^{-1}$. Furthermore, the term \texttt{"}mass\texttt{"} refers to the stellar mass unless explicitly mentioned.
	\section{Data and Method} \label{sec:method}
	\subsection{Data} \label{subsec:data}
	In this work, our primary objective is to explore the mass-based structural evolution of intermediate-mass SFGs using the data from the 3D-HST \citep{Brammer2012, Skelton2014} over five CANDELS high redshift fields: AEGIS, COSMOS, GOODS-N, GOODS-S, and UDS \citep{Grogin2011, Koekemoer2011}. In order to acquire structural parameters based on stellar mass maps, we matched our dataset with the stellar mass-weighted catalog generated by \cite{Mosleh2020}. By employing a pixel-by-pixel SED fitting technique on the 3D-HST imaging data, \cite{Mosleh2020} created a collection of resolved stellar mass maps and quantified the structural parameters of 5557 galaxies. Additionally, they constructed a catalog of approximately 3000 simulated galaxies to validate the accuracy of their methodology in measuring the mass-based parameters \citep[for more details, see][]{Mosleh2020}. Their analysis underscored that the sample achieved a 90\% mass completeness for SFGs with $\log(\mstar/\msun) \geqslant 9.8$ within the redshift range of 0.2 to 2.
	
	In addition, we used the light and mass-based parameters of the local galaxies from the SDSS dataset derived by \cite{Mosleh2017} as a reference sample at $z\sim0$ to pose further constraints on the estimated structural evolution of selected SFGs. \cite{Mosleh2017} calculated the structural parameters of about 1000 random local galaxies selected from the SDSS DR7 images at $0.03<z<0.06$ \citep{Abazajian2009}. They measured the one-dimensional light profiles by fitting a single \ser profile to the two-dimensional surface brightness of each galaxy using the \verb|GALFIT| \citep{Peng2010}. Then, by the SED fitting at each radius, they obtained the radial stellar mass and converted the light profiles to the mass profiles to estimate the light and mass-based parameters like the \ser index and half-mass radius.
	\subsection{Sample Selection} \label{subsec:selection}
	The first step in examining the structural evolution of intermediate to high-mass SFGs is to select galaxies that resemble the progenitors of these SFGs at each redshift. This requires that the stellar mass of the descendants should be determined in advance.  Theoretically, it has been found that star formation reaches the maximum efficiency around a halo mass of $10^{12}\msun$, which corresponds to a stellar mass of $10^{10.6}\msun$. This pivot mass might represent the transition point between the in situ SF-dominated to the merger-dominated regime \citep{Dekel2006, Mowla2019, Kaw2021}. Therefore, we were motivated to study SFGs with the current stellar mass ($M_{\ast,0}$) around this pivot mass: two samples with the $\log(M_{\ast,0}/\msun) = 10.3$ and 10.5, plus two more massive samples with $\log(M_{\ast,0}/\msun) = 10.7$ and 11. Next, it is possible to trace the main progenitors of a galaxy with a given final mass by estimating its SMGH. In this work, we obtained the SMGHs from two methods: the main-sequence integration (MSI) algorithm developed by \cite{Renzini2009} and \cite{Leitner2011}, which is referred to as Method I, and the average SMGH of selected disk-dominated SFGs from the IllustrisTNG (TNG50) simulations \citep{Nelson2019, Pillepich2019}, hereafter called Method II. 
	
	In the MSI method, we suppose that if the SFGs had relatively steady and in situ star formation histories (SFHs), they evolve along the observed star-forming main sequence at each redshift. According to this method, if the local SFGs have always been star-forming in the past, at each time step their stellar mass has been accumulated by in-situ star formation and recycled to the ISM by mass loss (ML). Therefore, by integrating the rate of mass growth (SF - ML) over time steps backward in time from the look-back time $t_0 = t(z = 0)$, it becomes possible to reconstruct the SMGH and SFH of a galaxy using its present-day stellar mass ($\mstar(t_0)$):
	\begin{equation} \label{eq:1}
		\mstar(t) = \mstar(t_0) - \int_{t_0}^{t} \left( \psi(\mstar,t) - \Re \right) \,dt 
	\end{equation} 
	Here, $\psi(\mstar,t)$ is the instantaneous SFR, and $\Re$ is the mass loss rate (MLR) defined as:
	\begin{equation} \label{eq:2}
		\Re(t) = \int_{t_0}^{t} \Phi(t\arcmin) \dot{f}_{ml}(t - t\arcmin) \,dt\arcmin 
	\end{equation}
	where $\Phi(t)$ is the SFH of this galaxy given by:
	\begin{equation} \label{eq:3}
		\Phi(t) = \psi(\mstar(t),z),
	\end{equation}
	and $\dot{f}_{ml}$ is the fractional mass loss rate which is quantified by \cite{Leitner2011} as: 
	\begin{equation} \label{eq:4}
		f_{ml} = C_0 \ln (\frac{t_2 - t_1}{\lambda} + 1) 
	\end{equation} 
	where $C_0 = 0.046$ and $\lambda = 2.76 \times 10^5$ yr for a simple stellar population with \cite{Chabrier2003} IMF. Since SFH is required to estimate $\Re$, but reconstruction of $\Phi(t)$ relies on knowing MLR, an iterative approach was presented by \cite{Leitner2011} to solve this problem. In this method, by assuming an initial value of $\Re=0.45$, equation~\ref{eq:1} is solved over each time interval. This allows us to estimate the SFR and mass of a galaxy step by step until reaching around its formation time. Subsequently, by employing the derived SFH, we can reassess the $\Re$ at each step. This process is reiterated multiple times to ensure that the results converge and are reliable.
	
	To estimate the instantaneous SFR of galaxies, we adapted the $SFR-\mstar$ relation of \cite{Speagle2014}, which is calculated based on the compilation of 64 MS relations from 25 studies at $z = [0, 6]$: 
	\begin{equation} \label{eq:5}
		\log \psi(\mstar,t) = (0.84 - 0.026 \times t)\log \mstar - (6.51 - 0.11 \times t) 
	\end{equation}
	where $t$ is the age of the universe (in Gyr).
	
	\begin{figure*}[ht!]
		\includegraphics[width=\textwidth]{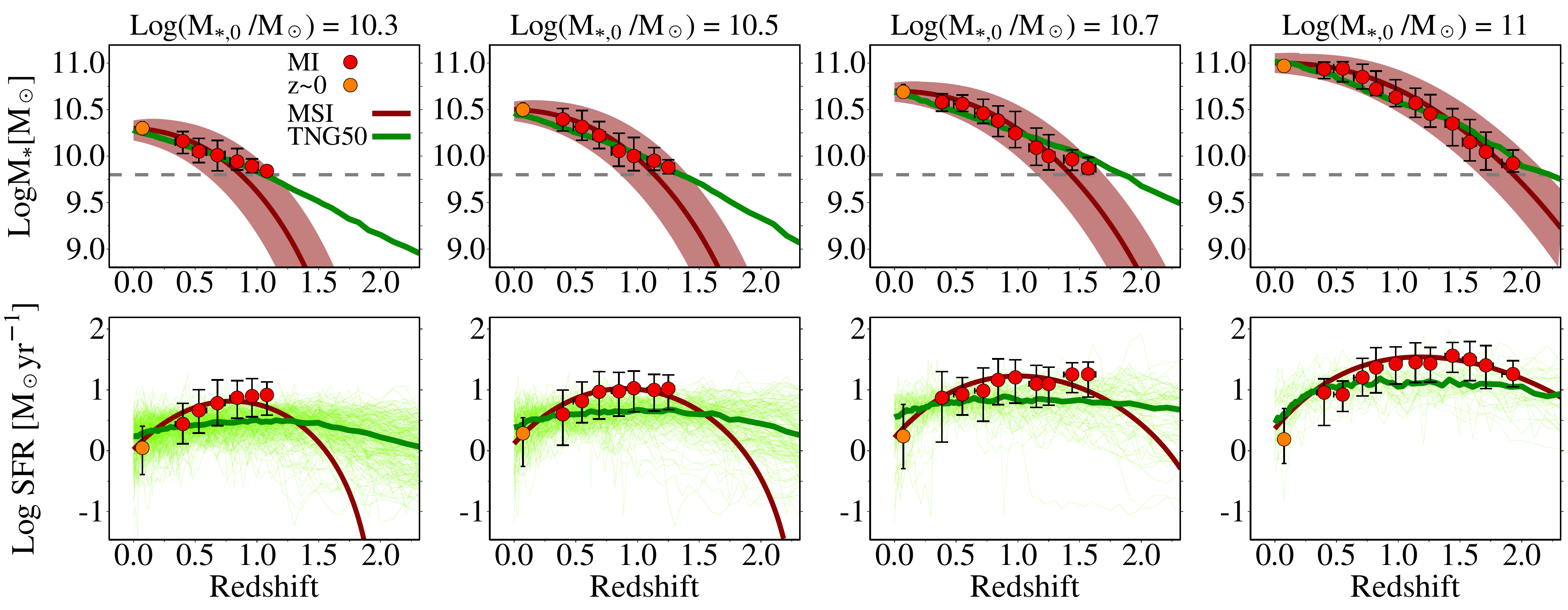}
		\caption{From left to right, mass growth (up) and star formation (bottom) histories of SFGs with a final stellar mass of $M_{\ast,0} = 10^{10.3}$, $10^{10.5}$, $10^{10.7}$ and $10^{11} \msun$ estimated by MSI method (dark red lines) using the SFR–mass relation of \cite{Speagle2014} and those selected from the IllustrisTNG (TNG50) simulations (thin light-green lines with the median in thick green lines). The dark red-shaded areas map the thresholds utilized to choose the samples based on SMGHs. The red circles are the median values of selected galaxies from the MSI method (Method I), whereas the orange circles represent the sample of local SFGs provided by \cite{Mosleh2017} using the SDSS DR7 images. In addition, the dashed gray lines visually illustrate the stellar mass limit of our primary sample, from the catalog of \cite{Mosleh2020}.}
		\label{fig1}
	\end{figure*}
	We selected the galaxies as progenitors whose mass and redshift reside respectively within the $\pm0.1$ dex and $\pm0.2$ of derived stellar mass assemblies to account for the uncertainties in estimating the stellar mass of observed galaxies and errors in calculating SMGHs. We adopted the $U - V$ versus $V - J$ \citep[hereafter UVJ;][]{Williams2009} rest-frame color criterion given in \cite{Mosleh2017} to distinguish SFGs from QGs. The galaxies with unreliable parameters ($\sim16\%$, see Section \ref{subsec:measurement}) were eliminated. Our final sample consists of 778, 1229, 1262, and 906 galaxies as possible progenitors of SFGs with the present-day stellar mass of $10^{10.3}, 10^{10.5}, 10^{10.7}$ and $10^{11} \msun$ using Method I over a redshift range of $z = [0.3 - 2]$. 
	
	For a reference sample at $z\sim0$, we chose the SDSS local SFGs with a stellar mass similar to the final mass of studied galaxies ($M_{\ast,0}$) in order to obtain more accurate fits. To do this, we picked SFGs from the sample of \cite{Mosleh2017} with the final mass of $\log(\mstar/\msun) = \log(M_{\ast,0}/\msun)\pm0.1$ dex. By applying this threshold, 146, 114, 83, and 28 local galaxies are chosen as local SFGs with $\log(\mstar/\msun)=10.3-11$. In Figure~\ref{fig1}, the red circles illustrate the median of stellar mass and SFR of progenitors derived from method I over $0.3<z<2.0$, whereas the local SFGs are shown in orange circles. The error bars represent the 16th to 84th percentiles of the distribution for each sample of galaxies. As can be seen, there is a slight difference in the estimated stellar mass between MSI and the observed median values from Method I for $\log(\mstar/\msun)<10$. Given that our observational data has a mass limit of $\log(\mstar/\msun) = 9.8$, by imposing the mass and redshift thresholds of $\pm0.1$ dex and $\pm0.2$, the median mass of the selected samples may exceed the MSI estimation at higher redshifts.
	
	To ensure that sample selection based on the MSI mass growth in which the merger effect is neglected, does not significantly alter our findings and also for a more accurate comparison with simulations, we attempted to obtain the mass assembly histories of SFGs using the SMGH calculated by the IllustrisTNG simulations. This paper uses the results of the smallest volume and highest resolution version of the IllustrisTNG Project, TNG50-1 \citep{Nelson2019, Pillepich2019}. To select the local SFGs with a specific final mass, the following thresholds were applied to galaxies at $z = 0$ (snapshot 99):
	\begin{itemize}
		\item Stellar mass within the radius of 30 kpc \citep{Engler2021}: $\log(\mstar/\msun) = \log(M_{\ast,0}/\msun)\pm0.15$ dex;
		\item Disk-to-total ratio obtained by classification 1 \citep[see][]{Du2020} for stars within 30 kpc: $D/T>0.5$;
		\item Star-formation rate (SFR) within a sphere of radius 30 kpc averaged across the last 100 Myr \citep{Donnari2019, Pillepich2019}: ${SFR} (\msun/{yr}^{-1})=[0.1, 10.0]$.
	\end{itemize}
	With these thresholds, 209, 142, 77, and 37 galaxies were selected in TNG50-1 as counterparts of disk-dominated SFGs with the final mass of $10^{10.3}, 10^{10.5}, 10^{10.7}$ and $10^{11} \msun$. We then used the \verb|SubLink| merger tree of the IllustrisTNG simulation \citep{Rodriguez-Gomez2015} to link the (sub)haloes at different snapshots and extract the main-progenitor branch for individual galaxies. The full assembly history of each parameter for SFG with a given present-day mass is derived by calculating the median values from a sample of entire progenitors of selected galaxies with the same mass at $z=0$. The obtained median mass is defined as the estimated SMGH from the TNG simulation, represented by the green line in Figure~\ref{fig1}. Using these SMGHs and applying the previous criteria, we selected 877, 1145, 1059, and 806 observed galaxies as the probable progenitors of Method II.
	
	As seen in Figure~\ref{fig1}, the TNG simulations predict a slower mass growth rate than the MSI method, especially for masses less than $\log(\mstar/\msun) \sim 9.9$. This discrepancy may result from not considering the merger or inaccurate mass loss or star formation rates in the MSI. However, as the minimum mass of our sample is $10^{9.8} \msun$, we can argue that the SMGHs obtained from both methods do not differ significantly within the data range of this study. In addition, the lower SFR values observed in TNG50 suggest different SMGHs may arise from the MSI initial assumption, which may not consistently hold, as SFGs may have gone through quenching periods during their evolution. To explore the impact of such events on mass accumulation and SFR, we identified quenched galaxies within the TNG simulation. By applying the sSFR thresholds of $\mathrm{sSFR}(z) < 1/(6\times t_H(z))$ \citep[as defined by][]{Tacchella2019}, we found that, on average, approximately 10\% of present-day SFGs in our TNG samples underwent a phase of quenching throughout their lifetime. Thus, experiencing a quenching period during the evolution can also contribute to the dissimilarity between the predicted SMGHs of the MSI method and TNG simulations. However, this difference might not be notable, at least when considering the galaxies within the low- to intermediate-mass range.
	
	Given the similar SFHs for simulated samples with and without ones with quenching periods, it appears that the lower SFR values in TNG50 compared to the MSI and observations could not be attributed to the existence of quenching phases during the evolution of SFGs. A similar discrepancy in the normalization values of MS between TNG and observational data was also reported by \cite{Donnari2019}. They found that, while TNG simulations have marginally higher SFR at $z=0$, their predictions are $\sim0.2-0.5$ dex lower than observational measurements \citep[including][]{Speagle2014} at intermediate redshifts.
	\subsection{Size Measurement} \label{subsec:measurement} 
	The mass-based sizes provided by \cite{Mosleh2020} are circularized, which is determined from the semi-major axis ($r_{e,SMA}$) and the axis ratio ($b/a$) by $r_{e,cir} = r_{e,SMA}\sqrt{b/a}$, to remove the effects of ellipticity. However, for SFGs or disk galaxies, the semi-major axis sizes of the ellipse that contains half of the total fluxes (total masses in this work) are often used \citep[see e.g.,][]{vanderWel2014}. Therefore, to adopt the standard measurement of half-mass radius along the semi-major axis and to assess the evolution of size and compare with previous studies, the size measurements were redone using 1D stellar mass surface density profiles derived from the stellar mass maps provided by \cite{Mosleh2020}. This approach reduces the effects of uncertainties in the mass maps, particularly in the outer regions of the galaxies. The procedure is nearly similar to the first size measurement method used in \cite{Mosleh2020} which followed the \cite{Maltby2018}. 
	
	In short, we first need to find the best-ﬁt \ser models from their 1D stellar mass surface density proﬁles. The 1D profiles were obtained using the \verb|IRAF| task \verb|ELLIPSE| \citep{Jedrzejewski1987}. However, to ensure consistency in the semi-major axis size $r_{e,SMA}$ with other studies, this process must be conducted in fixed-parameter mode, where the axis ratio ($b/a$) and position angle ($PA$) are held constant during the \verb|ELLIPSE| fitting procedure. Finding the best fixed-parameters from the stellar mass maps is more challenging. To address this, we utilized the output from the free-parameter mode \citep[i.e., $b/a$, $PA$, and the circularized effective radius $r_{e,cir}$ from][]{Mosleh2020} to estimate the appropriate $b/a$ and $PA$ values for the fixed-parameter mode. We calculated the average $b/a$ values from the radial profiles within a radius range of 2 times $r_{e,cir}$ to the radius where the stellar mass radial density profile reaches $10^7 \msun \mathrm{kpc}^{-2}$, beyond which background noise becomes dominant. We followed \cite{Mosleh2020} and used the simulated galaxies to ensure the method could recover the true values. A slight correction was necessary to implement after comparing the true and measured values of the $b/a$, using the $H_{160}$-band mock galaxies. The corrected average $b/a_{avg}$ method is then used and applied to obtain the average $PA_{avg}$. This method is also tested for the simulated mass maps.
	
	After determining the average position angle ($PA_{avg}$) and axis ratio ($b/a_{avg}$), and fixing the center, we obtained the stellar mass radial density profile using a fixed-parameter mode. In the next step, we generated a library of 1D \ser models for finding the best-fitting ones. Specifically, we used \verb|GALFIT| \citep{Peng2010} to first create 2D models on a grid spanning the size-\ser parameter space. The decision to generate 2D \ser models first, and then derive the library of 1D profiles, was driven by the two key considerations. First, this approach allows us to account for the two-dimensional distribution of the point spread function (PSF), which can significantly affect the profiles. By fitting ellipses to the PSF-convolved 2D models (with fixed center, $PA$, and ellipticity), we ensure that the resulting 1D profiles accurately reflect the impact of the PSF, thereby providing a more accurate comparison with observed data. Second, generating 1D profiles from 2D models ensures consistency with the observed data. Both the model and observed profiles undergo the same ellipse-fitting process, which minimizes potential biases that could arise if the profiles were derived using different methods. This consistency is crucial for the reliability of the subsequent chi-square minimization step, where we compare the normalized 1D mass density profiles of the galaxies with all normalized model profiles in the library to find the best-fitting model based on the minimum $\chi^2$ values.
	
	The robustness of the measurements was examined by utilizing simulated galaxies. Based on the results from the simulated mass maps, we found that on average the \ser indices were under-estimated by about $30\%$, while the $r_{e,SMA}$ at $z\gtrsim 1.2$ were over-estimated by $25\%$. Therefore, we have corrected these parameters, based on the outputs of these simulations.
	
	\begin{figure*}[ht!]
		\centering
		\includegraphics[width=0.9\textwidth]{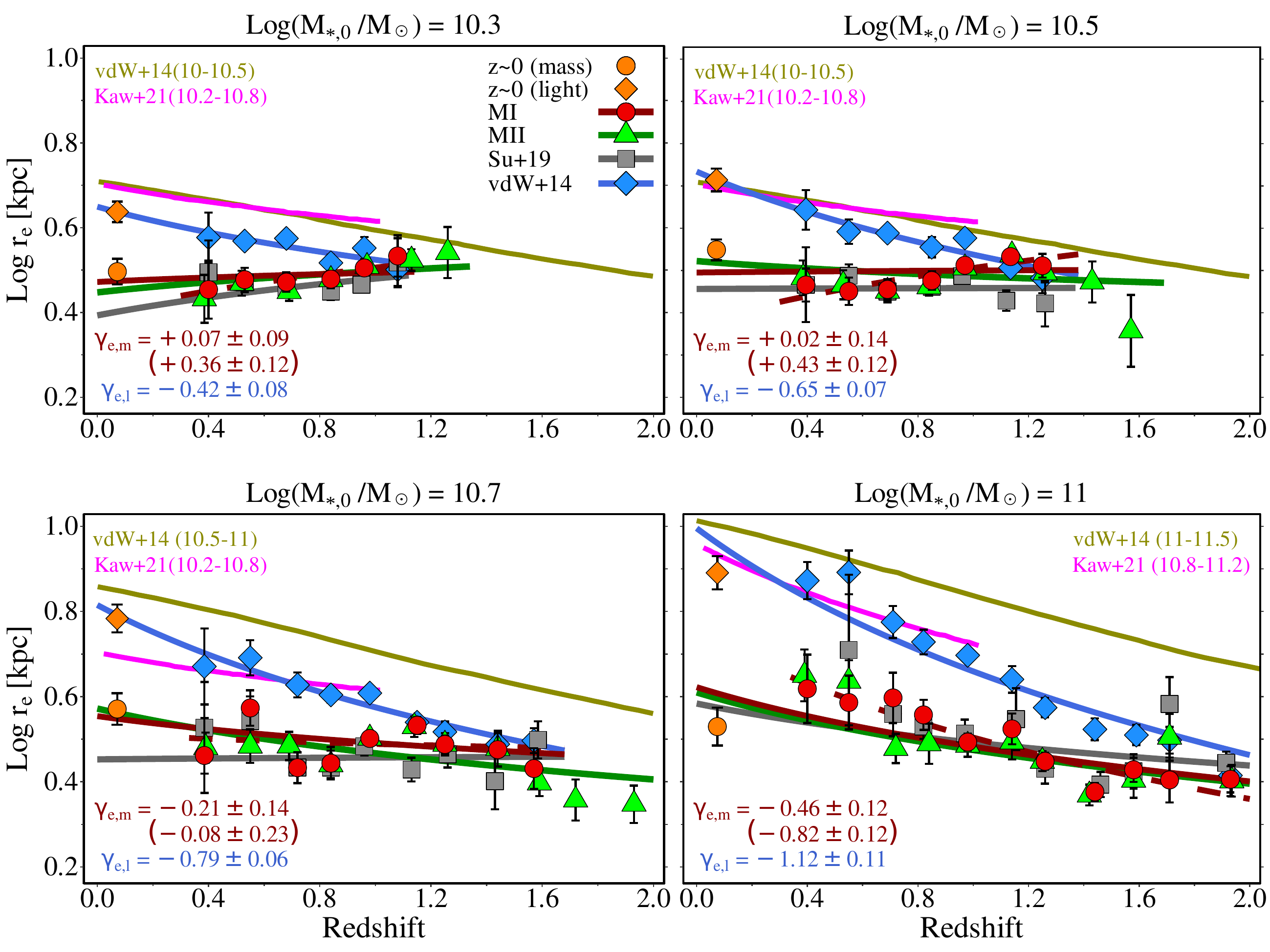}
		\caption{Redshift evolution of effective radius $r_{e,SMA}$ for progenitors of the SFGs at different stellar mass bins. The dark red and green lines represent the best fit to the medians of the form $r\propto(1+z)^\gamma$ inferred from Method I (red circles) and Method II (green triangles), while the dark red dashed lines show the fit to the median size evolution of Method I without considering the sizes at $z\sim0$. The half-mass and half-light radii of the local SDSS samples, obtained from \cite{Mosleh2017}, are shown by orange circles and diamonds, respectively. The median sizes (best-fit) estimated by \cite{Suess2019a} and \cite{vanderWel2014} are shown in gray squares (gray line) and blue diamonds (blue line), respectively. The best-fit slopes derived from Method I (with and without the local data) and light-based sizes of \cite{vanderWel2014} are denoted in dark red (without and with parenthesis) and blue, respectively. As seen, the mass-based radii \citep{Suess2019a} are in good agreement with our sizes, while $r_{e,light}$ \citep{vanderWel2014} is greater than $r_{e,mass}$, especially at higher redshifts. For comparison, the size evolution obtained by \cite{vanderWel2014} (olive line) and \cite{Kaw2021} (magenta line) for SFGs at fixed masses are over-plotted.}
		\label{fig2}
	\end{figure*}
	
	\begin{figure*}[ht!]
		\includegraphics[width=\textwidth]{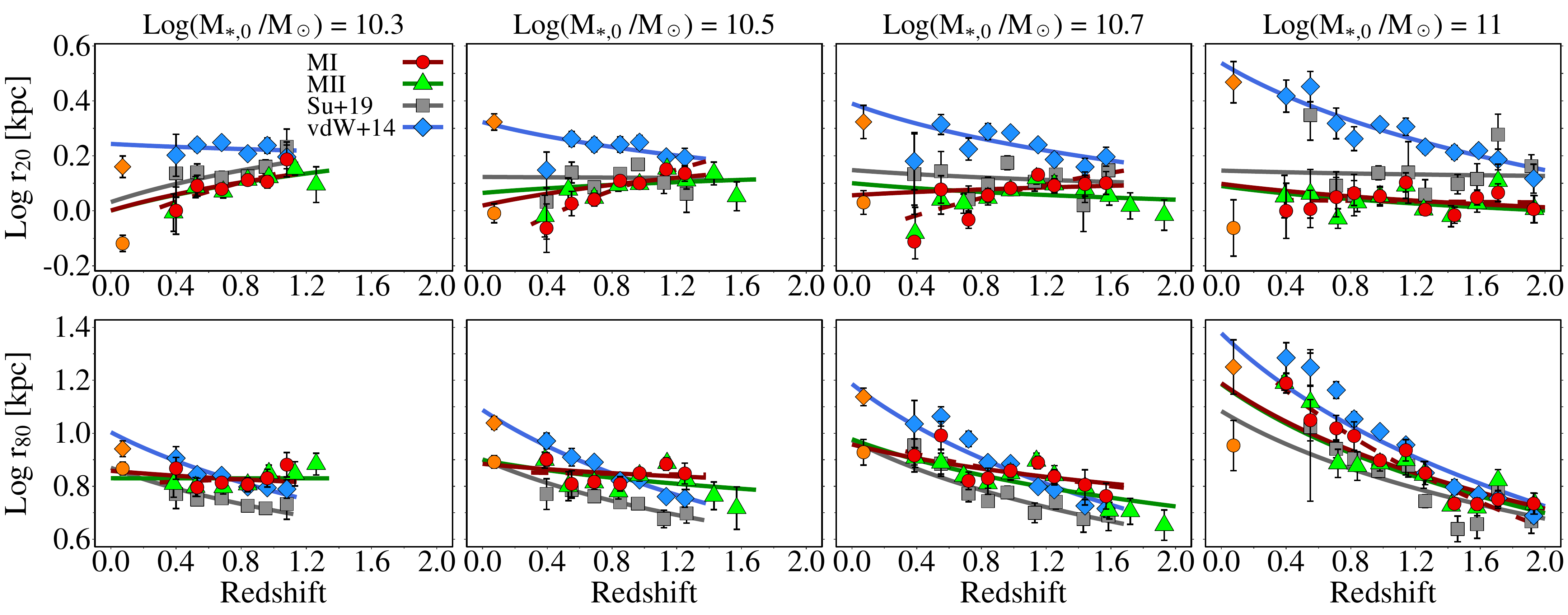}
		\caption{Evolution in the median of $r_{20}$ (top) and $r_{80}$ (bottom) of SFGs, same as Figure~\ref{fig2}. The $r_{80}$ sizes of the progenitors of higher mass SFGs have grown more than those with lower final mass, while their $r_{20}$ remains relatively constant. In contrast, the $r_{20}$ of smaller mass samples decreases with time, compared to the most massive SFGs, which have almost constant $r_{20}$ sizes. These results differ from the $r_{20}$ and $r_{80}$ estimated from the size catalogs of \cite{Suess2019a} and \cite{vanderWel2014}, which is probably related to the difference in \ser index and $r_{e,light}$ \citep[for][]{vanderWel2014} (see text for more details).}
		\label{fig3}
	\end{figure*}
	\begin{deluxetable*}{CCCCCC}[ht!]
		\tablenum{1}
		\tablecaption{Slopes of the median size evolution ($\gamma$) for different radii}
		\label{Table1}
		\tablewidth{0pt}
		\tablehead{
			\colhead{$\log(M_{\ast,0}/\msun)$} & \colhead{Size} & \colhead{Method I (w/o $z\sim0$)} & \colhead{Method II} & \colhead{Sizes from Su+19} & \colhead{Sizes from vdW+14}}
		\startdata & $r_{20}$ & $0.40\pm0.14~(0.64\pm0.23)$ & $0.39\pm0.12$ & $0.45\pm0.14$ & $-0.07\pm0.07$ \\
		$10.3$ & $r_{e}$ & $0.07\pm0.09~(0.36\pm0.12)$ & $0.17\pm0.12$ & $0.29\pm0.36$ & $-0.42\pm0.08$ \\
		& $r_{80}$ & $-0.13\pm0.13~(0.09\pm0.27)$ & $-0.01\pm0.13$ & $-0.54\pm0.07$ & $-0.74\pm0.06$ \\
		\\
		& $r_{20}$ & $0.30\pm0.24~(0.87\pm0.14)$ & $0.11\pm0.15$ & $-0.01\pm0.16$ & $-0.35\pm0.11$ \\
		$10.5$ & $r_{e}$ & $0.02\pm0.14~(0.43\pm0.12)$ & $-0.12\pm0.13$ & $0.01\pm0.29$ & $-0.65\pm0.07$ \\
		& $r_{80}$ & $-0.14\pm0.13~(-0.03\pm0.23)$ & $-0.26\pm0.13$ & $-0.61\pm0.06$ & $-0.94\pm0.04$ \\
		\\
		& $r_{20}$ & $0.08\pm0.22~(0.58\pm0.22)$ & $-0.13\pm0.14$ & $-0.10\pm0.14$ & $-0.50\pm0.10$ \\
		$10.7$ & $r_{e}$ & $-0.21\pm0.14~(-0.08\pm0.22)$ & $-0.35\pm0.11$ & $0.02\pm0.25$ & $-0.79\pm0.06$ \\
		& $r_{80}$ & $-0.36\pm0.14~(-0.45\pm0.22)$ & $-0.53\pm0.12$ & $-0.74\pm0.12$ & $-1.10\pm0.09$ \\
		\\
		& $r_{20} $ & $-0.18\pm0.12~(-0.03\pm0.15)$ & $-0.19\pm0.14$ & $-0.04\pm0.22$ & $-0.82\pm0.07$ \\
		$11.0$ & $r_{e}$ & $-0.46\pm0.12~(-0.82\pm0.12)$ & $-0.45\pm0.14$ & $-0.36\pm0.29$ & $-1.12\pm0.11$ \\
		& $r_{80}$ & $-0.99\pm0.21~(-1.57\pm0.13)$ & $-1.02\pm0.24$ & $-0.85\pm0.19$ & $-1.37\pm0.17$ \\
		\enddata
	\end{deluxetable*}
	\section{Results} \label{sec:result} 
	In this section, we examine how the structure of intermediate-mass SFGs changes with redshift, by analyzing the evolution of structural parameters such as size, \ser index ($n$), the axis ratio ($b/a$), and the stellar mass surface density within a radius of 1 kpc ($\Sigma_1$). We also compare radial profiles of stellar mass and sSFR of our samples.
	\subsection{Size Growth} \label{subsec:size}
	The effective radius ($r_{e,SMA}$) evolution of progenitors with $\log(M_{\ast,0}/\msun)= 10.3 - 11$ is shown in Figure~\ref{fig2}. Dark red and green lines represent the fit of the form $r\propto(1+z)^\gamma$ to the median size inferred from Methods I and II, respectively. To fulfill this, we calculated the fit on the median of samples selected using Method I (red circles) and Method II (green triangles) in 11 redshift bins from $z = 2.0$ to $z = 0.3$ and the median of SDSS local SFGs \citep[orange circles from][]{Mosleh2017}. The best fit to the median data obtained from Method I, with and without considering the SDSS data are shown in solid and dashed dark red lines, respectively. The uncertainty in each bin is derived from bootstrap resampling.
	
	According to Figure~\ref{fig2}, from $z = 2.0$ to $z = 0$, the progenitors of massive galaxies ($\log(M_{\ast,0}/\msun) = 11.0$) have a relatively slow evolution in size with $\gamma_{MI} = -0.46\pm0.12$ and $\gamma_{MII} = -0.45\pm0.14$, while the growth rate for both methods is much higher when local SDSS data is not considered. We can also observe a similar difference in growth rates of other parameters in our massive sample with and without considering $z = 0$, possibly due to the following reasons: Our SDSS sample has only 28 galaxies with $\log(M_{\ast,0}/\msun) = 11.0$, which leads to significant scatter in the median of structural parameters. The next problem is that the parameter measurement method differs in SDSS and 3D-HST/CANDELS data. For SDSS galaxies at a redshift of $z\sim 0$, the light profiles were first derived by fitting single \ser models to the observed two-dimensional surface brightness images in different filters using \verb|GALFIT|. From these best-fit models, the 1D light profiles are extracted for various filters, and these light profiles were then used to derive the deconvolved stellar mass profiles by performing SED fitting at each radial bin. Finally, the half-mass radius and other parameters were obtained from the stellar mass profiles \citep[see][for more details]{Mosleh2017}. Meanwhile, for galaxies in CANDELS fields, 2D stellar mass maps were calculated directly by fitting the best SED model for each pixel. After that, using the \verb|IRAF/ELLIPSE|, the stellar mass surface density profiles were estimated from the mass maps to derive the structural parameters. This leads to a slight mismatch between the parameters obtained by these two different methods and packages.
	
	For SFGs with $\log(M_{\ast,0}/\msun)= 10.7$, the effective radius does not significantly change over the redshift range of $0<z<2$ with $\gamma_{MI} = -0.21\pm0.14$ and $\gamma_{MII} = -0.35\pm0.11$. The median trend lines are reversed for galaxies with a mass below the pivot mass, although this size reduction is slightly noticeable when we exclude local galaxies in the analysis. In general, from $z\sim1.5$ to the current epoch, samples with $\log(M_{\ast,0}/\msun)\lesssim 10.7$ have a growth rate close to zero and hence, it is reasonable to conclude that their sizes remain almost constant throughout cosmic time. However, this trend marginally differs for SFGs with $\log(M_{\ast,0}/\msun)= 11$. We note that despite the approximately different slopes estimated with and without the local data, the median size evolution of galaxies follows a similar trend regarding the stellar mass. In addition, the results are consistent using Method II for selecting the progenitor and considering mergers in the SMGHs (green lines and best-fits values in Table~\ref{Table1}). Only progenitors of $\log(M_{\ast,0}/\msun)= 10.7$ at $z\sim1.5-2$ have smaller sizes and consequently, the effective radius increases slightly at this epoch.  
	
	To check our findings with other mass and light-based measurements, we cross-matched our data with size catalogs provided by \cite{Suess2019a} and \cite{vanderWel2014}. The median evolution of the half-mass radius from \cite{Suess2019a} (gray squares), and the half-light size from \citep{vanderWel2014} (blue diamonds) are shown as gray and blue trend lines in Figure~\ref{fig2}. Our half-mass radii almost agree with those estimated by \cite{Suess2019a}, except for some intervals, where there is considerable uncertainty in the median due to the small number of selected galaxies. The difference in these bins, as well as the lack of parameters for local galaxies, has caused the estimated slope for data from \cite{Suess2019a} to be slightly different from what we found, particularly for the two lower mass samples. Nevertheless, the general trend does not alter concerning our size measurements, and the results are consistent. However, the $r_{e,light}$ of \cite{vanderWel2014} is larger than $r_{e,mass}$, especially at low redshifts, which is in line with previous works suggesting that the average half-mass radii of galaxies are smaller than their half-light radii \citep{Szomoru2013, Mosleh2017, Suess2019b, Mosleh2020, Miller2023, vanderWel2024}. As a result, the growth rates of light-based sizes become steeper, which is more notable in high-mass samples. 
	
	We also plotted the evolution of half-light sizes from \cite{vanderWel2014} (olive lines) for SFGs at a fixed mass of $\log(\mstar/\msun)=$ 10 - 10.5 ($-0.52\pm0.08$), 10.5 - 11 ($-0.72\pm0.09$) and 11 - 11.5 ($-0.80\pm0.18$), and \cite{Kaw2021} (magenta lines) with $\log(\mstar/\msun)=$ 10.2-10.8 ($-0.29\pm0.01$) and 10.8 - 11.2 ($-0.77\pm0.02$). It can be seen from Figure~\ref{fig2} that the sizes of galaxies at fixed mass are larger than those inferred from the sample selected as progenitors of a given galaxy, even if we use the same data \citep[half-light radii adopted from][]{vanderWel2014}.
	
	\begin{figure*}[ht!]
		\includegraphics[width=\textwidth]{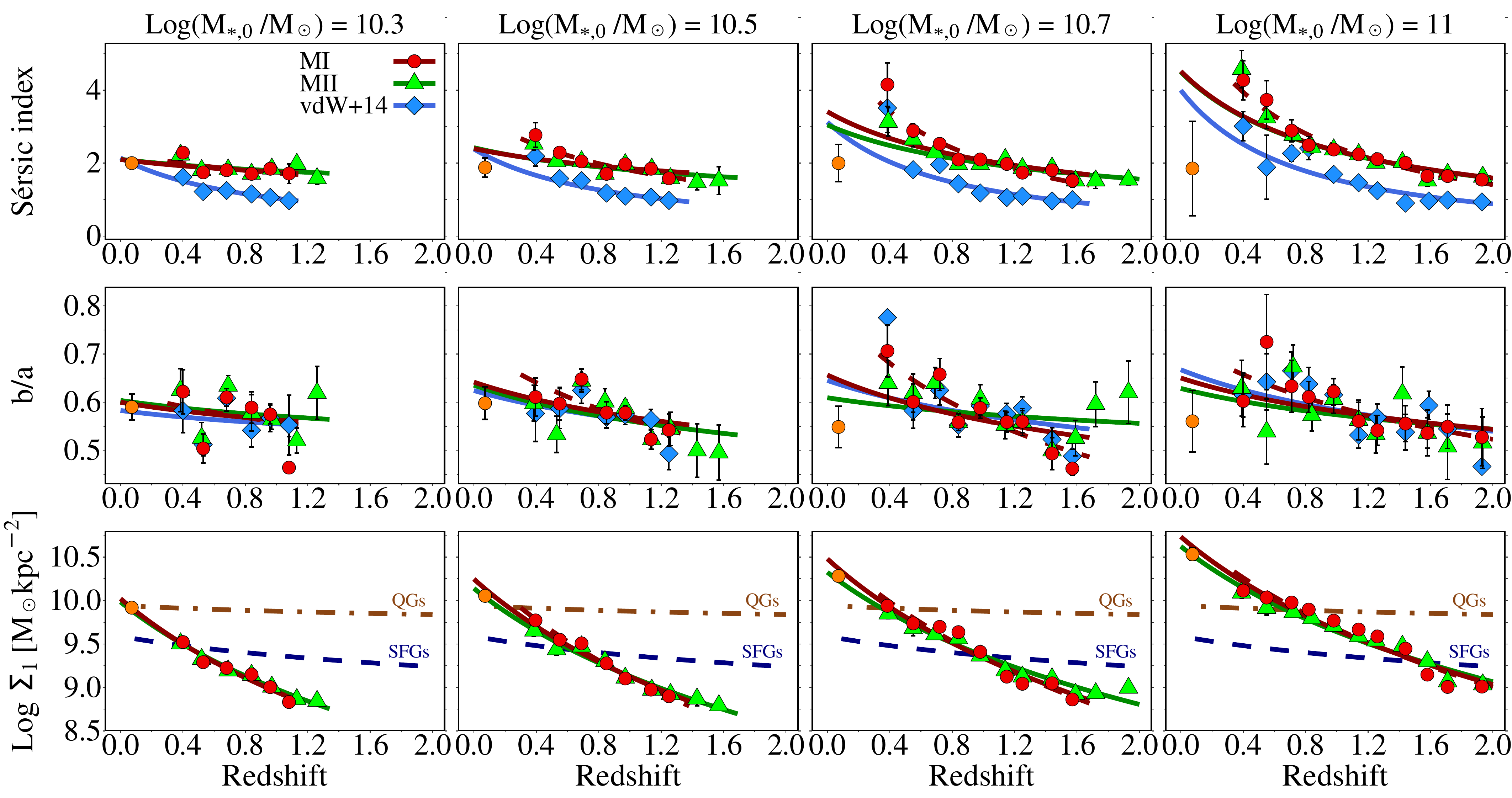}
		\caption{From top to bottom, redshift evolution of \ser index ($n$), axis ratio ($b/a$), and central stellar-mass density ($\Sigma_1$), same as Figure~\ref{fig2}. The dashed navy and dash-dotted brown lines in the bottom panels represent the fit to all SFGs and QGs in our primary sample. It should be noted that \cite{vanderWel2014} fitted single \ser models to the light profiles to measure the parameters; therefore, the estimated $n$ is smaller than the ones found using the mass profiles.}
		\label{fig4}
	\end{figure*}
	Moreover, we applied additional measurements of $r_{20}$ and $r_{80}$, representing the radii that contain $20\%$ and $80\%$ of a galaxy's total stellar mass, respectively. The evolution in the median $r_{20}$ and $r_{80}$ of our samples is shown in the top and bottom panels of Figure~\ref{fig3}. Although the slope of size growth of Method I decreases when considering the local samples, in general, the evolution of Method I is consistent with those derived by Method II within the errors. The $r_{20}$ size of massive galaxies slightly grows since $z\sim2.0$, which can, therefore, be inferred that the mass fraction inside these regions remains approximately constant. In contrast, the $r_{20}$ radii of other samples decrease toward low redshift, especially for $z<1.2$. It can be concluded that the stellar mass grows faster in the central parts of these galaxies, which leads to the $r_{20}$ radius reaching about 1 kpc in  $z\sim0.3$.
	
	Finally, using $r_{80}$, the size of massive SFGs grows toward low redshifts, which leads to the largest $r_{80}$ for SFGs with $\log(M_{\ast,0}/\msun) = 11$ at later epochs. For lower mass samples, this rate slows down so that for the progenitors with $\log(M_{\ast,0}/\msun) = 10.5$, it can be seen that the slope is closely flat for both methods. These trends between the stellar mass of galaxies and their sizes seem to indicate that the mass growth in the massive galaxies is relatively faster in the outskirts, unlike the low-mass SFGs, in which mass builds up at a higher rate in the interior regions. For intermediate mass galaxies with masses around the pivot mass, mass growth rates in the inner and the outer parts are relatively comparable; therefore, these SFGs evolve self-similarly, as inferred from the evolution of half-mass radius.
	
	For comparison with the mass and light-based sizes of \cite{Suess2019a} and \cite{vanderWel2014}, we tried to estimate $r_{20}$ and $r_{80}$ from $n$ and $r_e$ of their catalogs using Equation (3) of \cite{Miller2019}. It is important to note that the \ser indices from the \cite{vanderWel2014} catalog, adopted by \cite{Suess2019a} as well, were derived from the light profiles and are smaller than those measured by \cite{Mosleh2020} using mass profiles. Therefore, the $r_{20}$ radii obtained from the structural parameters of \cite{Suess2019a} and \cite{vanderWel2014} are larger than what we found. By contrast, since $r_{80}$ is proportional to $n$, the $r_{80}$ sizes from the parameters of the \cite{Suess2019a} catalog are smaller than our measurements. In comparison, the $r_{80}$ radii of \cite{vanderWel2014} are approximately consistent with our sizes because of the larger values of half-light radii than half-mass radii. 
	\subsection{Morphological Evolution of Progenitors} \label{subsec:sersic}
	\begin{deluxetable*}{CCCCC}[ht!]
		\tablenum{2}
		\tablecaption{The best-fit parameters for the structural evolution of samples same as Table \ref{Table1}}
		\label{Table2}
		\tablewidth{0pt}
		\tablehead{
			\colhead{$\log(M_{\ast,0}/\msun$)} & \colhead{Parameter} & \colhead{Method I (w/o $z\sim0$)} & \colhead{Method II} & \colhead{Data from vdW+14}}
		\startdata & $n$ & $-0.26\pm0.17~(-0.40\pm0.31)$ & $-0.23\pm0.14$ & $-1.05\pm0.09$ \\
		$10.3$ & $b/a$ & $-0.09\pm0.16~(-0.16\pm0.34) $ & $-0.08\pm0.12$ & $-0.07\pm0.11$ \\
		& $\Sigma_1$ & $-3.56\pm0.22~(-3.52\pm0.42) $ & $-3.33\pm0.14$ & -- \\
		\\
		& $n$ & $-0.37\pm0.24~(-0.91\pm0.25)$ & $-0.42\pm0.17$ & $-1.08\pm0.21$ \\
		$10.5$ & $b/a$ & $-0.17\pm0.10~(-0.35\pm0.14)$ & $-0.18\pm0.10$ & $-0.15\pm0.09$ \\
		& $\Sigma_1$ & $-3.73\pm0.25~(-4.31\pm0.25)$ & $-3.36\pm0.16$ & -- \\
		\\
		& $n$ & $-0.72\pm0.27~(-1.39\pm0.18)$ & $-0.61\pm0.15$ & $-1.28\pm0.31$ \\
		$10.7$ & $b/a$ & $-0.22\pm0.13~(-0.52\pm0.11)$ & $-0.08\pm0.08$ & $-0.17\pm0.13$ \\
		& $4\Sigma_1$ & $-3.79\pm0.27~(-4.13\pm0.36)$ & $-3.19\pm0.18$ & -- \\
		\\
		& $n $ & $-0.95\pm0.20~(-1.35\pm0.10)$ & $-0.95\pm0.21$ & $-1.38\pm0.244$ \\
		$11.0$ & $b/a$ & $-0.16\pm0.08~(-0.30\pm0.08)$ & $-0.13\pm0.09$ & $-0.19\pm0.09$ \\
		& $\Sigma_1$ & $-3.56\pm0.27~(-3.80\pm0.34)$ & $-3.26\pm0.18$ & -- \\
		\enddata
	\end{deluxetable*}
	In the top panel of Figure~\ref{fig4}, the evolution of the median \ser index ($n$) is illustrated for progenitors tracked via Method I (with and without $z\sim0$) and II, as well as those cross-matched with light profile’s \ser indices measured by \cite{vanderWel2014} (see Table~\ref{Table2} for the values). The $z\sim0$ samples have different \ser indices, in particular for the massive SFGs. As previously mentioned, this disparity stems from the employing dissimilar methodologies for measuring the parameters and the statistically small number of galaxies in these bins. The \ser index from the \cite{vanderWel2014} catalog exhibits lower values but a steeper slope. 
	
	\begin{figure*}[ht!]
		\centering
		\includegraphics[width=0.9\textwidth]{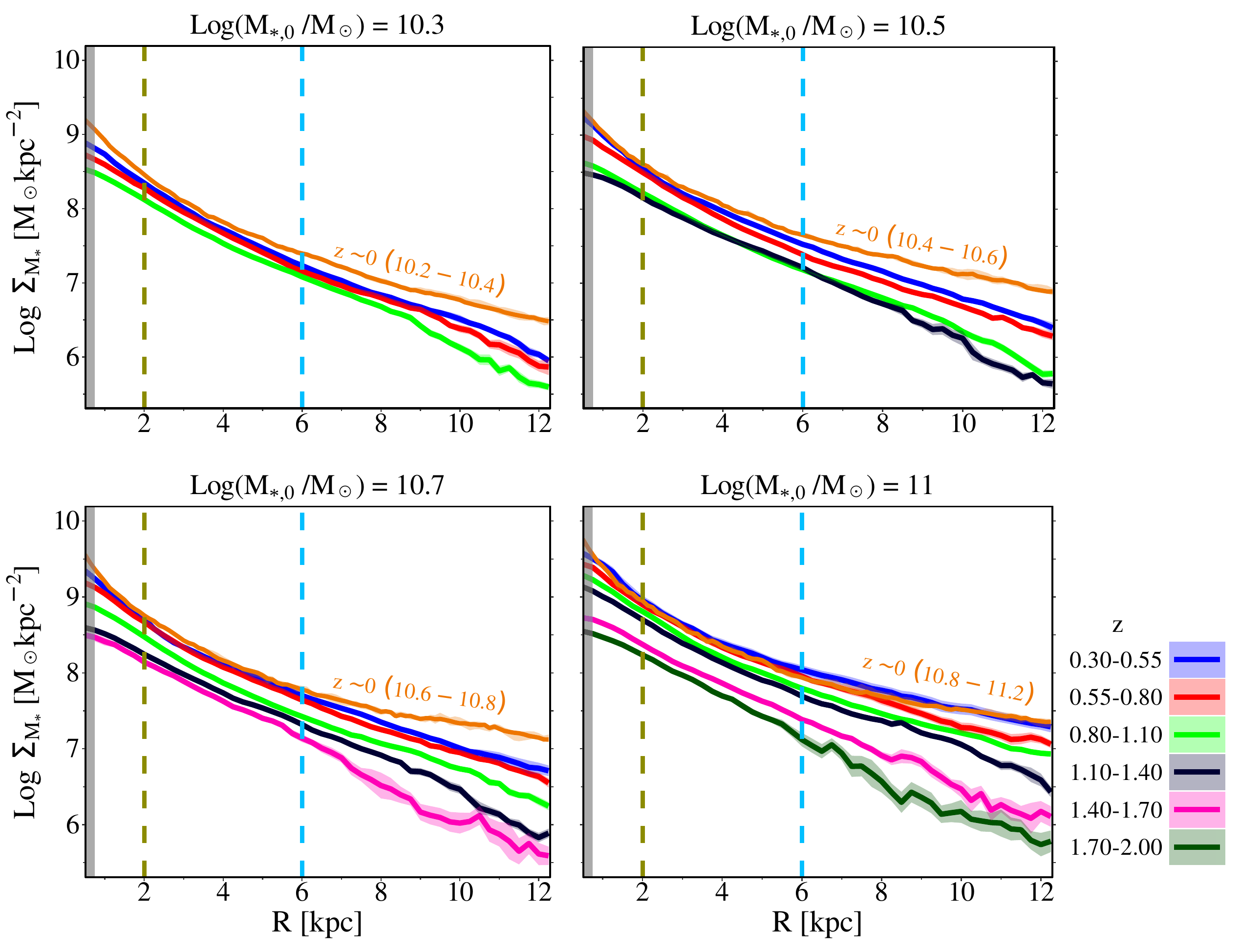}
		\caption{Stellar-mass surface density profiles of SFGs with $\log(M_{\ast,0}/\msun) = 10.3$, $10.5$, $10.7$ and $11$ in different redshifts, along with the profiles of the local SFGs at $z\sim0$ calculated by \cite{Mosleh2017} (dark orange lines) using SDSS DR7 images. The dashed olive and cyan lines illustrate the radius of 2 and 6 kpc. The slopes of profiles in the middle part ($2 \mathrm{kpc} < R < 6 \mathrm{kpc}$) evolve at nearly the same rate in all redshifts, whereas toward the lower redshift, slopes of the inner ($R < 2 \mathrm{kpc}$) and outer regions ($6 \mathrm{kpc} < R < 12 \mathrm{kpc}$) increase and decrease, respectively.}
		\label{fig5}
	\end{figure*}
	In addition to $n$, we used the redshift evolution of central surface mass density within 1 kpc ($\Sigma_1$) to investigate the build-up of central structures. In the bottom panels of Figure~\ref{fig4}, we display the median evolution of $\Sigma_1$ as a function of redshift for progenitors selected by Method I and II and all SFGs and QGs in our mass-based data from $z \sim 0.0$ to 2.0. As can be seen, median central densities increase significantly over the redshift range of our study. Generally, both methods have relatively consistent median slopes within their errors, even without considering data at $z \sim 0$. However, the redshift evolution of $\Sigma_1$ is generally in agreement with the average increase in the \ser index, suggesting that the interior regions of SFGs become more concentrated and these galaxies form a dense central core (bulge or pseudo-bulge). The central density of our two more massive samples reaches the average values of QGs at $z\sim0.4$ and $z\sim0.8$, respectively, suggesting that the rate of evolution of a galaxy's central structure probably correlates with its stellar mass. At these epochs, the \ser index also increased and surpassed $n = 3$, comparable to those of QGs, which agrees with the results of \cite{Barro2017}. They found that the SFGs whose $\Sigma_1$ overlapped with QGs had \ser indices close to the fully quenched galaxies and exhibited a similar morphology.
	
	Finally, we used the minor-to-major axis ratio ($b/a$) to study the morphological changes of our SFGs and examine the results obtained from the evolution of $n$ and $\Sigma_1$. The axis ratio slightly increases with cosmic time and reaches the values of 0.6 and 0.7 for samples below and above the pivoting mass in the local universe, which are close to the mean apparent axis ratio of disk- and bulge-dominated galaxies calculated by \cite{VincentRyden2005}. Using SDSS DR3 data, \cite{VincentRyden2005} studied a sample of about 96,000 galaxies and measured the mean apparent axis ratio of 0.59 and 0.74 for galaxies with pure exponential ($n\lesssim1.2$) and de Vaucouleurs profiles ($n\gtrsim3.3$), respectively.
	\subsection{Radial Stellar Mass Assembly} \label{subsec:mass_density}
	Studying the stellar mass distribution within the galaxies and comparing the mass buildup in different radii could lead to a better understanding of the stellar mass assembly in galaxies. Thus, we extend our analysis to use the stellar mass surface density profiles and track the redshift evolution of radial mass distribution for each sample in a non-parametric way. Figure~\ref{fig5} presents our samples' median mass density profiles within a radius of 12 kpc. To provide a more clear depiction of the profiles, they are presented in 6 bins.  For comparison, we adopted the stellar mass density profiles of four samples of local SFGs with stellar mass $\log(\mstar/\msun)=10.2-11.2$, provided by \cite{Mosleh2017} using the SDSS DR7 images at $0.06<z<0.08$ (dark orange lines). Olive and cyan vertical dashed lines represent the radii of 2 and 6 kpc, and the gray shaded area shows the maximum HWHM of PSFs. The majority of profiles exhibit little to no change in their general shape with redshift, and their slope remains approximately constant, suggesting that stellar mass is built up at all radii. Although, for high-mass samples, stellar mass grows slightly faster in the outer parts ($6\mathrm{kpc} < R < 12\mathrm{kpc}$) at high redshifts. The slope of mass density profiles tends to be shallower in these regions at $z < 1.1$ for samples with $\log(M_{\ast,0}/\msun)\geq10.7$. In contrast, inner mass profiles ($R < 2\mathrm{kpc}$) become marginally steeper, which is consistent with the concentration of the central region.
	\begin{figure*}[ht!]
		\centering
		\includegraphics[width=0.9\textwidth]{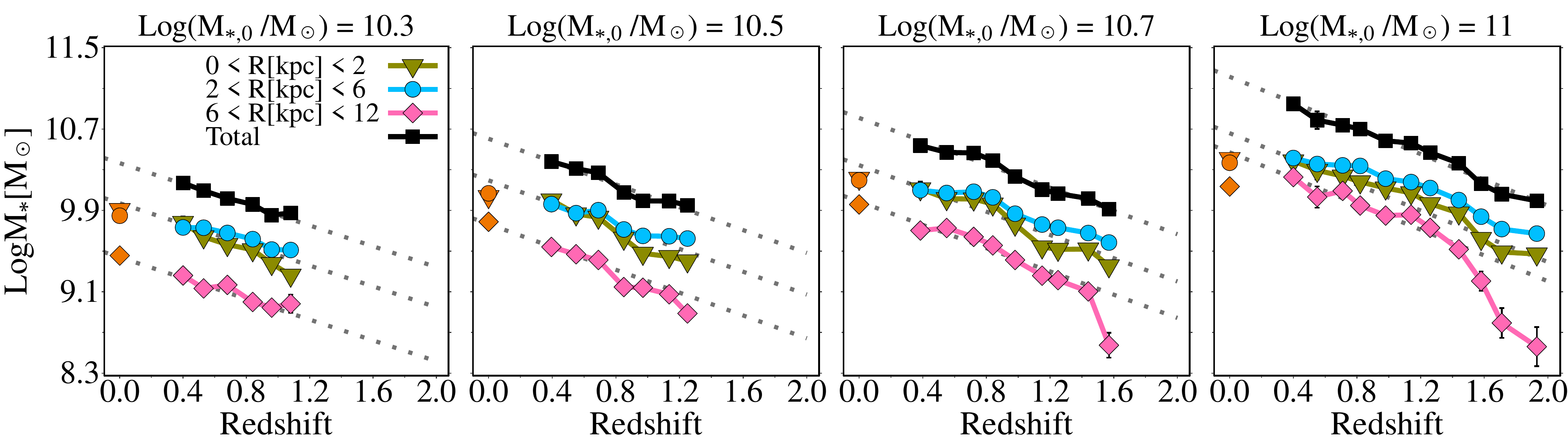}
		\caption{Mass assembly in the inner (olive triangles), middle (cyan circles), and outer (pink diamonds) regions of progenitors. The dark orange points represent the stellar mass of the corresponding regions at $z\sim0$, calculated from the profiles of the local SFGs derived by \cite{Mosleh2017}. The black squares and dotted lines illustrate the total stellar mass over redshift bins and the corresponding best-fits. The mass growth rate of the different regions is in line with the size and $\Sigma_1$ evolution of progenitors.}
		\label{fig6}
	\end{figure*}
	
	We further quantified and compared the mass build-up in the different areas along with the mass density profiles. In Figure~\ref{fig6}, olive triangles, cyan circles, and pink diamonds display the stellar mass assembled in the inner ($R < 2\mathrm{kpc}$), middle ($2\mathrm{kpc} < R < 6\mathrm{kpc}$), and outer regions ($6\mathrm{kpc} < R < 12\mathrm{kpc}$), while the black squares give the total stellar mass. At $z\sim0.4$, the stellar masses within each sample's inner and middle regions have nearly the same values. This is also the same for the local galaxies, so it is reasonable to suppose that the mass distributed inside 2 kpc and 2 - 6 kpc is almost the same at lower redshifts. For a more accurate analysis of evolutionary rates in different regions, we have plotted the linear fit to the total mass as dotted lines in Figure~\ref{fig6}. It is evident that the slopes of the mass assembly differ between the central and outer parts in each mass regime. The mass evolution in the inner radii of galaxies with $\log(M_{\ast,0}/\msun)= 10.3$ is somewhat faster, with 0.52 dex growth in the mass inside $R = 2$ kpc, while the stellar mass added to the outer regions increased only by 0.29 dex, which confirms our findings about the size evolution of these low mass SFGs. Also, the central region of more massive SFGs experienced a noticeable mass increase (0.47 dex and 0.63 dex) between $z = 1.1-0.7$ and $z = 1.7-1.0$ for progenitors with $\log(M_{\ast,0}/\msun)= 10.7$ and 11, respectively. The rapid mass accumulation within the core of these SFGs is consistent with the accelerated growth of their central densities ($\Sigma_1$) at these epochs. In addition, the outskirts of these galaxies had an accelerated mass growth phase at high redshifts. The stellar mass within the range $6\mathrm{kpc} < R < 12\mathrm{kpc}$ of SFGs with $\log(M_{\ast,0}/\msun)= 10.7$ and 11 increased by 0.53 and 1.31 dex at $1.45<z<1.65$ and $1.1<z<2.0$. 
	\begin{figure*}[ht!]
		\centering
		\includegraphics[width=0.9\textwidth]{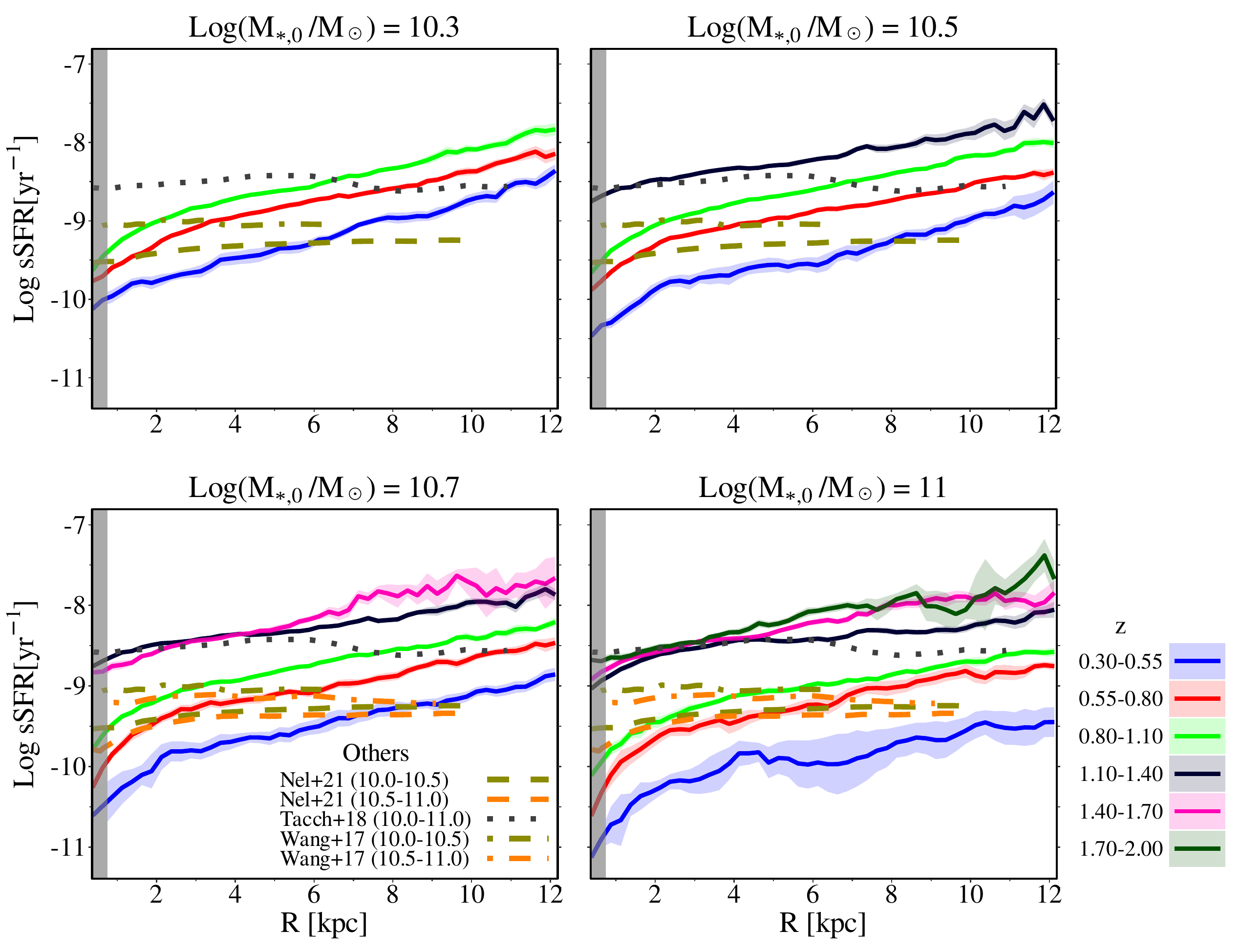}
		\caption{Comparison of our sSFR profiles (solid lines) with dust-corrected profiles measured by \cite{Wang2017} (dash-dotted lines) for SFGs in stellar mass ranges of $\log(\mstar/\msun) = [10-10.5]$ (olive), and $[10.5-11]$ (orange) at $1.0 < z < 1.2$, and \cite{Tacchella2018} (dotted gray lines) for star-forming main sequence galaxies with a stellar mass between $10^{10} \msun$ and $10^{11} \msun$ at $z\sim2.2$. The dashed lines represent the sSFR profiles of galaxies with $\log(\mstar/\msun) = [10.0,10.5]$ (olive), and $[10.5-11.0]$ (orange) at $z\sim1.0$ from TNG50 simulation derived by \cite{Nelson2021}.}
		\label{fig7}
	\end{figure*}	
	\subsection{sSFR profiles} \label{subsec:SFR}
	Using the resolved maps derived from the pixel-by-pixel SED fitting, we estimated the star-formation rate (SFR) and, accordingly, the specific SFR (sSFR) profiles. Figure~\ref{fig7} shows the median sSFR profiles within the 12 kpc radius for 6 redshift bins. The shaded regions indicate bootstrapped scattering of the median SFR. These profiles show that the maximum star formation activities occur at $z > 1.2$. While the sSFR decreases over time, the decline is faster in the inner regions compared to the outer parts until $z\sim0.4$. It occurs more rapidly for high-mass galaxies, leading to shallower radial profiles at lower redshifts and higher masses. 
	
	In addition, it is interesting that higher mass samples exhibit a centrally suppressed star formation such that the sSFR in the central regions of SFGs with $\log(M_{\ast,0}/\msun) = 11$ attains the quenching threshold of $10^{-11} yr^{-1}$. These observations are in line with the rapid growth of the \ser indices and $\Sigma_1$, indicating that in high-mass SFGs, a central dense component (bulge or pseudo-bulge) with very low SFR has been formed and is growing. The mass-dependent star formation activity in the central region of galaxies agrees with previous works \citep{Tacchella2015, Tacchella2018, Morselli2019}.
	\section{Discussion} \label{sec:discussion}
	\begin{figure*}[ht!]
		\centering
		\includegraphics[width=0.9\textwidth]{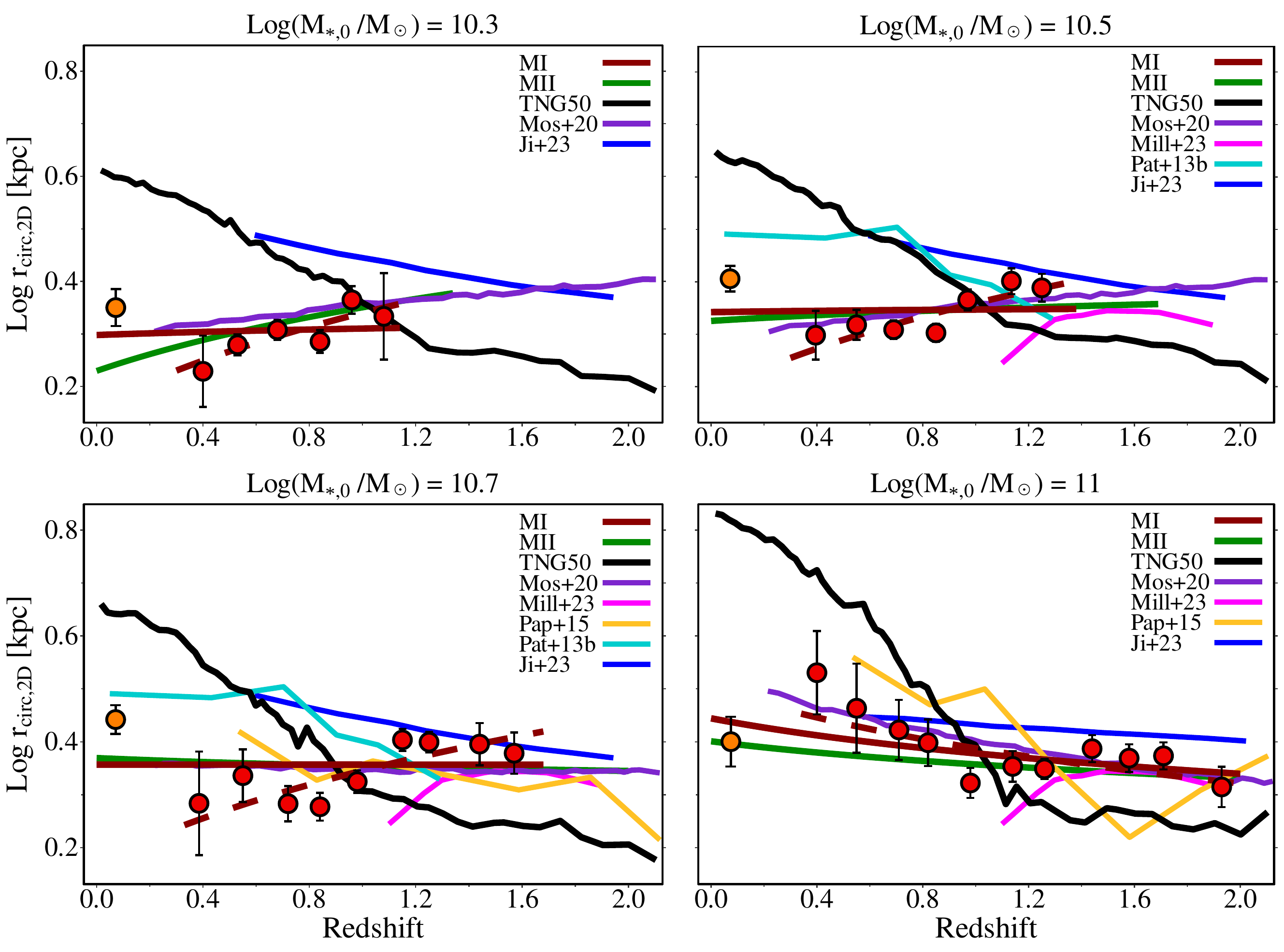}
		\caption{Comparison of the circularized size evolution of SFGs obtained from the first method with (dark red lines) and without (orange lines) considering the $z\sim0$ SDSS data with other works. It can be seen that the mass-based sizes of \cite{Mosleh2020} (purple lines) and \cite{Miller2023} (magenta lines) change very little, while the half-light radii \cite{Ji2023} (blue lines), \cite{Patel2013b} (cyan lines), and \cite{Papovich2015} (orange lines) are often larger at lower redshifts.}
		\label{fig8}
	\end{figure*}
	In this section, we first compare our findings to those from previous simulations and observational studies and investigate the possible factors contributing to the variation in size growth rates. Then we examine the evolution of the surface density of mass and SFR, as well as the spin parameter of the halo. Finally, we discuss the implications of our results on our understanding of the history of galaxy growth on galaxy growth.
	\subsection{Comparison with observational studies} \label{subsec:comp_obs}
	The distinctive aspect of this work is the selection of possible progenitors of galaxies and employing mass-based structural parameters. Due to these features, our results challenge some earlier findings stating that the sizes of SFGs increase considerably toward lower redshifts. In this section, we compare the circularized radii, $r_{50}$, of our samples with those of other observational studies. Figure~\ref{fig8} illustrates the circularized size evolution of galaxies inferred from Method I and II, as well as those measured by previous studies \citep{Patel2013b, Papovich2015, Mosleh2020, Miller2023, Ji2023}. The slope of the size growth rate depends on the stellar mass for both methods, although the obtained slope is shallower than previous estimates. 
	
	Our $r_{50}$ measurements are consistent with the mass-based sizes of \cite{Mosleh2020} (purple lines) and \cite{Miller2023} (magenta lines), as can also be seen for effective radii in Figure~\ref{fig2}. \cite{Mosleh2020} measured the mass-based structural parameters of galaxies in fixed stellar mass bins over the redshift range $z=0.3-2.0$. Despite the same data, the slopes are slightly different because we identified the probable progenitors instead of the analysis at fixed mass. Our estimated size evolution trends are also in line with the slow size evolution of SFGs within the range of $10.5<\log(\mstar/\msun)<11$ observed by \cite{Miller2023}. They studied the evolution of the color gradient and half-mass radius of $\sim3000$ galaxies measured with the non-parametric framework of \verb|imcascade| \citep{Miller_vanDokkum2021} in 3D-HST/CANDELS fields. This mild evolution is evident in other studies investigating the evolution of mass-based/NIR sizes along the semi-major axis \citep[e.g.,][]{Suess2019b}. Based on the JWST imaging, \cite{Ward2024} also reported a shallower rate of evolution at the fixed mass of $10^{10.7}\msun$ ($0.63\pm0.07$ for SFGs at $0.5<z<5$) compared to previous studies such as \cite{vanderWel2014}.
	
	Although \cite{Papovich2015} used optical sizes, their estimated trends for the progenitor-selected samples are compatible with our sizes within the redshift range of this study. However, despite using the same method (MSI), our estimated sizes differ more from the results of \cite{Patel2013b} compared to \cite{Papovich2015}, who employed the abundance matching method. This could be attributed to the type of galaxies: \cite{Patel2013b} focused solely on SFGs, whereas \cite{Papovich2015} examined both SFGs and QGs. In addition, \cite{Patel2013b} assumed a constant M/L profile to convert light to mass profiles. Like \cite{Papovich2015}, \cite{Ji2023} utilized the sizes taken from the catalog of \cite{vanderWel2012} to analyze the size evolution of SFGs with $\log(\mstar/\msun)=10.3-10.8$ and $\log(\mstar/\msun)>10.8$ after eliminating the progenitor effect. Using the Prospector, they reconstructed the SFH of each galaxy to identify the probable progenitors with the same mass and SFR at the formation epoch. Their analysis revealed a much slower evolution for SFGs compared to \cite{Papovich2015}, which can be partly attributed to separating the sample into SFGs and QGs.
	
	\cite{Whitney2019} studied the evolution of Petrosian sizes measured by the non-parametric approach of CAS for two samples: a sample selected at the constant number density of $n=1\times10^{-4}Mpc^{-3}$ (corresponding to the present-day stellar mass of $\log(M_{\ast,0}/\msun)\sim11.25$) and a sample at the fixed mass range of $\log(\mstar/\msun)=9.0-10.5$. The number-density-selected sample grows at a rate of $-0.53\pm0.16$, which is in good agreement with the observed evolution of effective radius in our massive sample ($-0.46\pm0.12$). The other sample (at fixed mass) revealed steeper slopes ($-0.92\pm0.03$), supporting the idea that the rate of size evolution is faster in the fixed-mass-selected samples relative to the progenitor-selected ones. In addition, our estimated slopes for the evolution of $r_e$ are compatible with the measurements of \cite{George2024} for effective radii of SFGs at fixed mass, though, they found a more rapid evolution in the size of samples with evolving mass at $0.1<z<0.9$. This disparity may arise from the methods used to calculate the mass growth and estimate the size. To measure the SMGH up to $z=0.1$ for each galaxy with a specific mass at $z=0.9$, they considered only star formation without accounting for mass loss ($\Re$ in the equation of \ref{eq:2}), resulting in much faster mass growth. In addition, the sizes of galaxies with a given mass were estimated from the size-mass relation they observed at each redshift.
	\begin{figure*}[ht!]
		\centering
		\includegraphics[width=0.9\textwidth]{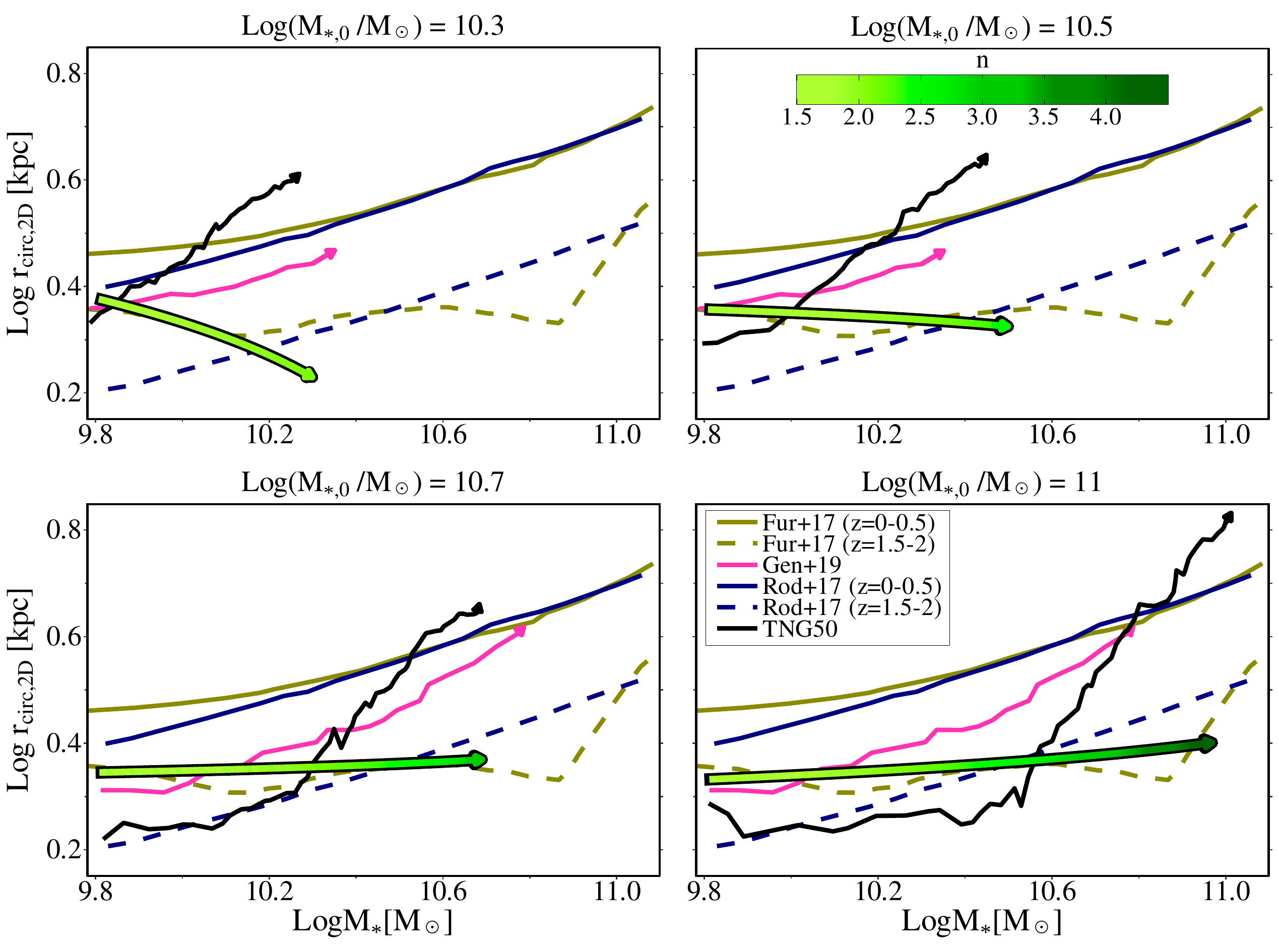}
		\caption{Half-mass radius-stellar mass evolutionary track of progenitors (thick green lines with arrows at the end) with four different final masses, color-coded account to their \ser indices. We utilized the two-dimensional circularized radii to compare our results with the simulations, including the evolutionary track of TNG100 (magenta lines) and the size-mass relation of $Ref-L100N1504$ EAGLE (olive lines) taken from \cite{Genel2018} and \cite{Furlong2017}. In addition, we directly compared the sizes from observation with simulation by extracting the half-mass radius of galaxies selected from the TNG50 (black lines) to calculate the SMGH. Generally, the growth rates predicted by the semi-empirical model of \cite{Rodriguez-Puebla2017} (navy lines) and simulations are faster than observations. This difference is even more significant if we compare the size growth predicted by the TNG50 with that observed from selected SFGs with similar mass at each redshift.}
		\label{fig9}
	\end{figure*}
	\subsection{Comparison with Simulations} \label{subsec:comp_sim}
	Despite numerous attempts to reproduce the evolution of galaxies, the simulated sizes do not agree well with observations \citep{Scannapieco2012, Snyder2017, Roper2022}. While simulations like IllustrisTNG and EAGLE have made progress in reducing this difference by fine-tuning parameters like feedback and dust attenuation, their predictions remain inconsistent with observed measurements. As discussed in Section \ref{sec:intro}, simulations exhibit a considerable increase in the sizes of SFGs, whereas the observed half-mass radii do not change significantly. Therefore, we attempted to examine this disparity by comparing the size evolution of SFGs in the simulations to those observed. To carefully analyze the predicted sizes, the 3D half-mass radii ($r_{3D}$) of the simulations are converted to projected circularized sizes ($r_{2D}$) using the analytical relation of $r_{2D}=r_{3D}/1.3$, derived by \cite{vandeVen2021}.
	
	In order to conduct a fair comparison with simulations, we extract the half-mass sizes of SFGs that we selected from the TNG50-1 as the progenitors of each sample with a specific mass. The evolution of median sizes taken from TNG50-1 is shown as black lines in Figure~\ref{fig8}. As can be seen, there is a notable disparity between the evolution of the projected half-mass sizes of TNG50 and our observations at $z<0.8$. Furthermore, we directly compared the half-mass radius from the stellar mass tracks of Method II (thick green lines with arrows at the end color-coded to \ser indices) with those predicted by the TNG50-1 simulations (black lines with arrows) in Figure~\ref{fig9}. As can be seen, there is a notable disparity between the projected half-mass sizes of TNG50 and our observations at later epochs. It should be noted that the evolutionary paths of progenitors chosen via Method I are similar to those obtained through Method II. Such a trend can also be seen for the central and satellite progenitors in the TNG100 measured by \cite{Genel2018}. The evolutionary path of these galaxies that remain star-forming at $z=0$ with final mass ranges of $\log(M_{\ast,0}/\msun)=10.3–10.45$ and $10.7–10.85$, is shown by pink lines with arrows in Figure~\ref{fig9}. To comprehend this trend, it is important to note that in the TNG simulations, the local size-mass relations estimated from the light profiles \citep[e.g.,][]{Shen2003} have been considered in order to discern between various simulated models.
	
	For a more comprehensive analysis of this discrepancy, we use the size-mass relation of the EAGLE simulations, determined by \cite{Furlong2017}, in two redshift intervals: 0.-0.5 (solid olive lines) and 1.5-2 (dashed olive lines). \cite{Furlong2017} examined the size evolution of active and passive galaxies from the EAGLE simulations over the range $0<z<2$ in two resolutions of 100 and 25 cMpc box: $Ref-L100N1504$, $Ref-L025N0752$, and $Recal-L025N0752$. While our evolutionary trajectories align with the predicted size-mass relations of $Ref-L100N1504$ EAGLE simulation at $z=1.5-2$, they diverge in the lowest redshift bin. Such dissimilarity between the simulations' half-mass sizes and observations was also reported by \cite{Suess2019b, Hasheminia2022} and \cite{Miller2023}. It is worth noting that our results show relative consistency (particularly in the higher mass range) with $Recal-L025N0752$, which underwent recalibration for stellar feedback and AGN parameters in the 25-cMpc box. Nevertheless, the significant size growth in simulations is also inconsistent with the mild evolution of the observed mass-based sizes at $z<1$ \citep{Suess2019b, Mosleh2020, Hasheminia2022}. 
	
	Finally, we analyze whether our findings align with the size-mass relation of a semi-empirical model developed by \cite{Rodriguez-Puebla2017} in two redshift ranges of $0<z<0.5$ (solid navy lines) and $1.5<z<2$ (dashed navy lines). They utilized a semi-empirical approach to establish a connection between galaxies and their host halos, allowing them to estimate the merger rates, SFHs, and structural parameters. To determine the size, they adopted the local size-mass relation of \cite{Mosleh2013} with the redshift-dependent coefficients constrained by the size-mass evolution of \cite{vanderWel2014}. As evident from Figure~\ref{fig9}, the evolution of their size-mass relation proposes a size growth with a steeper slope compared to our results. Adopting the mass-size relation of the local galaxies based on their light profiles can be one of the origins of the discrepancies.
	
	\begin{figure*}[ht!]
		\centering
		\includegraphics[width=\textwidth]{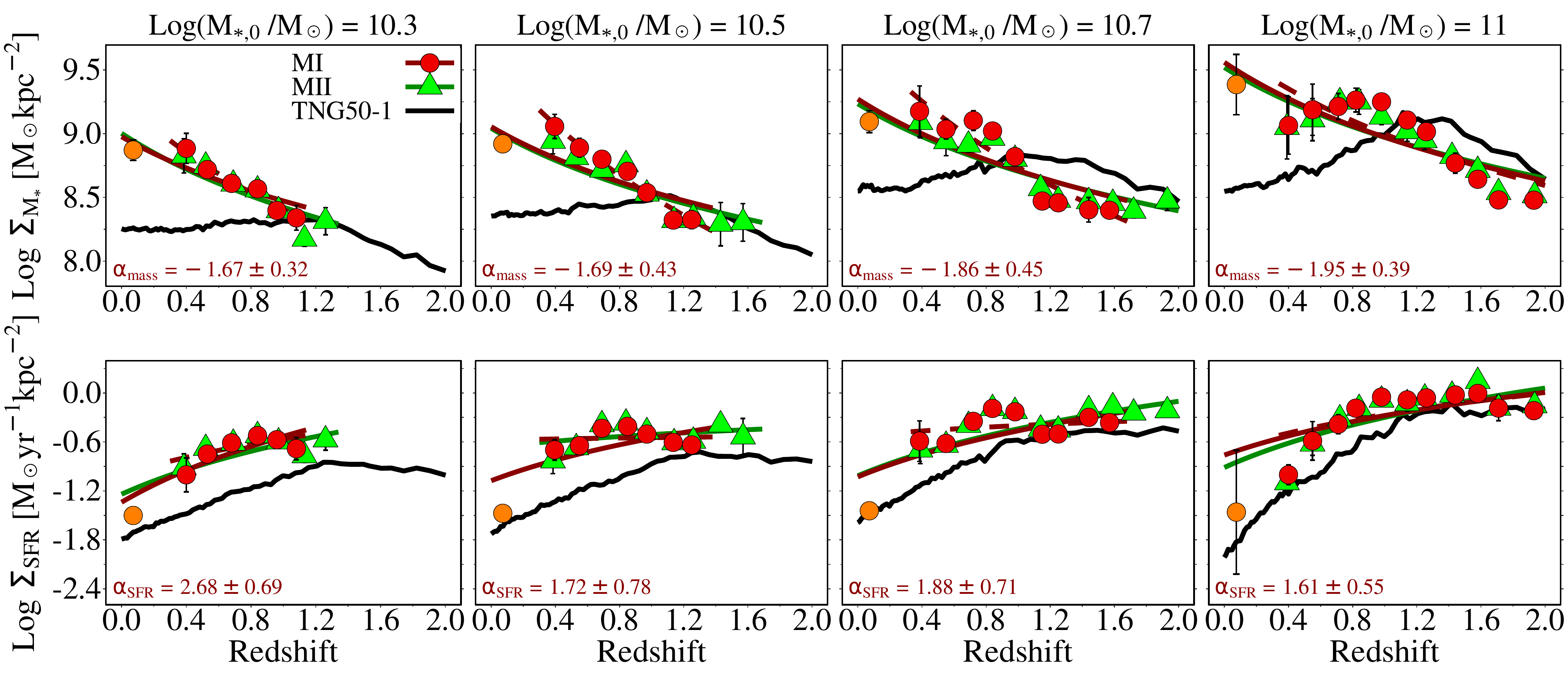}
		\caption{Stellar mass surface density ($\Sigma_M$), (top panel) and star formation surface density, $\Sigma_{SFR}$, (bottom panel) of selected SFGs within the half-mass radius. $\Sigma_{SFR}$ decreases toward low redshifts, in agreement with the TNG50 simulation (black lines) and previous observational works, whereas $\Sigma_M$ increases from $z\sim1.2$ in contrast to the simulation.}
		\label{fig10}
	\end{figure*}
	Differences in the size evolution rates can be reflected in quantities, such as the surface density of stellar mass ($\Sigma_M$) and SFR ($\Sigma_{SFR}$). Therefore, we compare the $\Sigma_M$ and $\Sigma_{SFR}$ of our samples with those predicted by TNG50 (black lines) in Figure~\ref{fig10}. The stellar mass surface density is defined as a function of the half-mass radius by:
	\begin{equation} \label{eq:6}
		\Sigma_{M} = \frac{\mstar}{2\pi r_{50}^2}
	\end{equation}
	The same approach is used to calculate SFR density.
	
	In general, the stellar mass surface densities of our observed samples tend to rise. More massive galaxies exhibit higher stellar mass surface density: for samples with $\log(M_{\ast,0}/\msun)=10.3$ and 11, the median $\Sigma_M$ values in the local universe reach $10^{8.9}$ and $10^{9.4} \msun/\mathrm{kpc}^2$, respectively. It seems that the mass surface density of massive galaxies reaches the highest value around $\log(\Sigma_M)\sim9.2$, and then remains relatively constant, but the lower mass SFGs keep building the stellar mass inside the effective radius. Although TNG50 SFGs show a similar mass dependence, their surface densities demonstrate a decreasing pattern at $z<1$. As Method II relies on TNG50's SMGHs for sample selection, this distinction between our results and TNG50 possibly arises from the notable growth of the half-mass radii in simulations since $z\sim1$.This rapid expansion indicates the necessity of some mechanisms in simulation to regulate the size growth.
	
	As shown in the bottom panel of Figure~\ref{fig10}, the SFR surface densities obtained from both methods (MI \& MII) are almost identical, indicating an overall decline in the $\Sigma_{SFR}$, especially at $z\sim0.8$. Our samples at $z > 0.6$ show a positive relation between their stellar mass and $\Sigma_{SFR}$, i.e., more massive SFGs tend to have a higher $\Sigma_{SFR}$ value. Our observational trends align with previous works \citep{Ono2013, Shibuya2015, Shibuya2019, Morishita2024}. In addition, the findings of \cite{Morishita2024} provided evidence that supports the correlation between SFR and the stellar mass of galaxies, while \cite{Shibuya2015} presented a contradictory conclusion. However, the values we obtained are within the range $-1.5<\log(\Sigma_{SFR})<0.15$, as derived by \cite{Shibuya2015}. Hence, the distinct conclusion reached by \cite{Shibuya2015} can be ascribed to selecting samples in fixed mass bins. 
	
	The SFR surface densities of TNG50 (black lines) are almost parallel to our findings due to the slower decline in their SF activities. As seen in Figure~\ref{fig1}, the assembled mass is nearly equal while the SFR in TNG50 is lower than our measurements. Hence, it can be inferred that the contribution of other channels, such as merger and accretion, to the stellar mass accumulation, is higher in TNG simulations compared to observations. Moreover, the decline in surface density of SFR could be related to the dynamical effects. According to the simple model of \cite{Lehnert2015} for sSFR evolution, increasing the angular momentum (AM) of the accreted gas toward lower redshifts can lead to a reduction in both the surface density of the gas and SFR. The model predicts that the high AM causes the transfer of accreted gas to larger radii and the accumulation of mass in the disk's outer regions, consequently leading to an expansion in size. Therefore, the decreasing trend of SFR surface density evolution and significant size growth in the simulation may be attributed to the high AM of accreted gas, as suggested by this model.
	
	To get more insights into the origins of dissimilarity in sizes, we discuss the following contributed mechanisms:\\
	\begin{itemize}
		\item \textbf{Feedbacks and Radial Gradient of M/L Ratio}: The impact of feedback mechanisms on the efficiency of star formation and scaling relations was examined by many studies \citep{Dutton2009, Aumer2013, Shankar2013, Noguchi2018}. For example, \cite{Dutton2009} which used a disk-galaxy evolution model. It has been found that feedback can lead to a shallower slope in the size-mass relation of galaxies. Thus, considering the relative agreement between $Recal-L025N0752$ and the observational data, as well as the findings of previous studies such as \cite{Dutton2009}, we emphasize that fine-tuning the feedback can enhance the consistency between observed and simulated sizes, as suggested by \cite{Suess2019b}. In addition, disagreement in sizes can also be attributed to dissimilarity in $M/L$ ratios between simulations and observations. The little disparity between the growth rates of the half-mass and half-light radii in different simulations indicates that the distribution of mass and light is relatively consistent in simulations. Moreover, it is evident from Figures 2 and 3 of \cite{Genel2018} that the $r_{e,mass}/r_{e,light}$ in TNG100 is larger than the observed ratios \citep{Mosleh2017, Suess2019a, Mosleh2020, Miller2023, vanderWel2024}. Thus, it can be concluded that the $M/L$ in TNG100 is higher than the observed ratio obtained by \cite{Mosleh2017} and \cite{Suess2019a}. One explanation for this difference could be the calibration of $M/L$ using local observables in simulations, without accounting for the color gradient's evolution with redshift, especially at $z<1.5$. However, for a more thorough investigation, it is crucial to measure the $M/L$ or color gradient in simulations and compare it with the findings from observational studies.\\
		\item \textbf{Redistribution of stellar mass}: Besides the mentioned possibilities, the dynamical effects can also change the stellar mass distribution in galaxies. Hence, comparing the dynamical evolution of SFGs in simulation and observation can help to clarify the impact of this factor on the size discrepancy. In the following, we investigate the redshift variation of AM in TNG simulations as a proxy for dynamical changes and compare it with our findings. 
		
		We analyze the evolution of `galaxy spin', $\lambda_{gal} = \lambda_{halo}~(j_{d}/m_{d})$, using the simple disk formation models \citep{Fall1980, Fall1983, Mo1998}. According to these models, baryons settled in disks will have a mass and AM that is a relatively constant fraction of the mass ($m_d$) and AM ($j_d$) of their host dark matter halo. These models predict the disk scale length is a fixed fraction of the halo size given by:
		\begin{equation} \label{eq:7}
			R_{gal} = \frac{\lambda_{halo}}{\sqrt{2}} \left( \frac{j_{d}}{m_{d}} \right) R_{halo}
		\end{equation} where,
		\begin{equation} \label{eq:8}
			R_{halo} = \left( \frac{G M_{halo}}{100 {H^2(z)}} \right) ^{1/3} = \frac{V_{halo}}{10 H(z)}
		\end{equation}
		Here, $\lambda_{halo}$ and $H(z)$ are the halo spin and the Hubble parameter at given redshift $z$. Therefore, the size of the disk evolves as $(1+z)^{-1}$ at a constant halo mass, while at constant circular velocity, it grows even faster with $(1+z)^{-1.5}$.
		By substituting Equation (\ref{eq:8}) into Equation (\ref{eq:7}), with $E=-M_{halo} V_{c}^2/2$ and $r_e=1.68 R_{gal}$, the galaxy spin could be related to the halo mass and effective radius of a disk galaxy via:
		\begin{equation} \label{eq:9}
			r_e \simeq 0.26 \lambda_{halo} \left( \frac{j_{d}}{m_{d}} \right) \left( \frac{G M_{halo}}{H(z)^2} \right)^{1/3}
		\end{equation}
		By comparing the size in the first to the last redshift bin, the stability of galaxy spin over time can be estimated as:
		\begin{equation} \label{eq:10}
			\frac{\lambda_{gal,0}}{\lambda_{gal,z}} = \left( \frac{M_{halo,z}}{M_{halo,0}} \right)^{1/3} \left( \frac{H(0)}{H(z)} \right)^{2/3} \frac{r_{e,0}}{r_{e,z}}
		\end{equation}
		
		By extracting the corresponding halo mass of each stellar mass from TNG50 and calculating the changes in the size of SFGs selected by Method II, we find that observed galaxies with $\log(M_{\ast,0}/\msun)=10.3$, 10.5, 10.7, and 11 have $\frac{\lambda_{gal,0}}{\lambda_{gal,z}}= 0.47$, 0.68, 0.59, and 0.46 since $z=1.38$, 1.75, 2, and 2, respectively. If the halo spin parameter remains constant over the redshift range of this study, our samples now possess around half their initial retention fraction of specific angular momentum (sAM), $j_d/m_d$. By assuming that the initial sAM of the baryons and the dark matter are the same, the average fraction of AM retained in the disk from the halo is close to 55\% within the past $10$ Gyr. This is relatively consistent with those measured by \cite{Swinbank2017}, who found that rotationally supported galaxies with $V/\sigma > 5$, retain $\sim70\%$ of their dark matter specific AM. Their investigation revealed that galaxies with a higher ratio of $V/\sigma > 5$ tend to have higher specific AM, lower SFR surface density, more pronounced bulges, and smoother disks at fixed mass. Furthermore, \cite{Burkert2016} conducted a study on the H$\alpha$ IFU kinematics of 359 SFGs at $z\sim0.8–2.6$ and stellar mass range of $\log(\mstar/\msun)\sim9.3–11.8$. They concluded that for SFGs 60\% to 90\% of sAM of their halo transferred to baryonic disk since $z=2.6$.
		
		According to the sizes extracted from TNG50, the simulations predict higher values of 1.11, 0.97, 0.94, and 1.30 for corresponding SFGs. Since $m_d$ in simulation and observation are relatively similar, we can define the retention fraction of sAM as the AM fraction of the disk to its host halo. Assuming a constant $\lambda_{halo}$ suggests that the fraction of AMs in the simulated galaxies is relatively stable, while observations demonstrate a substantial reduction in this ratio. This stability in the retention fraction of AM of simulated SFGs, which results in the redistribution of stellar mass and subsequent growth in size, could be attributed to the high efficiency of AM transfer from the accreted halo gas to the disks in simulations.
	\end{itemize}

	\subsection{Implications for the Evolution of SFGs} \label{subsec:implications}
	Our results show that how the mass assembles at different radii depends on the galaxy's stellar mass: within the redshift range of this study, for SFGs with $\log(M_{\ast,0}/\msun)=10.3$, the interior regions assemble relatively faster, while the outskirts of massive galaxies ($\log(M_{\ast,0}/\msun)=11$) are undergoing much higher rates of mass growth than the inner parts. We can also see this trend reflected in the evolution of the $r_{20}$ and $r_{80}$ sizes so that for the $r_{20}$ size, the slope of the evolution becomes shallower with increasing mass, while the growth rate of $r_{80}$ increases. In addition, $\Sigma_M$, $\Sigma_1$, and \ser index of SFGs are also correlated with their mass, so galaxies with higher mass tend to have higher mass surface density, more compact central component, and prominent bulge.
	
	However, the analysis of samples with $\log(M_{\ast,0}/\msun)\sim 10.5-10.7$ reveals that the mass growth rates are nearly comparable across all radii. This finding supports the notion that mass accumulates self-similarly in this mass regime, which is aligned with a lack of significant changes in the half-mass size of these SFGs. The modest evolution in size, accompanied by the considerable increase in mass, causes the surface density of these galaxies to rise constantly, reaching around $10^9\msun/\mathrm{kpc}^2$ in the local universe.
	
	These results suggest massive galaxies are undergoing a different evolutionary stage. Over the redshift range of $1.2 < z < 2.0$, the inner parts, experience only three times increase in mass, while the mass in the outer regions increases by a factor of 12. Following a period of spatially consistent mass build-up, the growth rate in the outskirts rises again since $z\sim0.7$, contrasting with a slow mass accumulation rate in the inner regions. During this epoch, the growth of $\Sigma_M$ and $\Sigma_1$ slows down, with the central surface density surpassing the median value of QGs. It can be interpreted as the initial stage of mass saturation \citep{Tacchella2015, Abdurrouf2023, Lapiner2023} in the interiors of massive SFGs. In addition, although all samples show a common trend of rising sSFR profiles, the decline in sSFR is more pronounced in the inner regions of massive galaxies. The sSFR in their central 1 kpc region drops from $10^{-8.8} yr^{-1}$ at $1.20 < z < 1.35$ to the quenching threshold $10^{-11} yr^{-1}$ at $0.45 < z < 0.60$, indicating the inside-out quenching. These findings support the idea that the star formation activities of galaxies cease when their central density reaches a critical value \citep{Fang2013, Barro2017, Whitaker2017}. 
	
	Recent studies suggested that when haloes reach a critical mass ($M_{shock}$) establish stable shocks that can heat the infalling gas to near the virial temperature, causing a hot circumgalactic medium. The cold streams cannot penetrate through this medium at $z < z_{crit}$, leading to the suppression of gas supply and, therefore, long-term quenching \citep{Birnboim2003, Keres2005, Dekel2006, Tacchella2016, Behroozi2019}. At $z < 1$, the halo mass of our massive sample surpasses the critical mass, suggesting that these galaxies may experience long-term quenching as cold flows have been halted since $z\sim2$.
	
	Although at $z < 1$ the massive SFGs have lower sSFR in outer regions compared to other samples, their sizes exhibit more rapid growth. A possible explanation could be the accretion or merger events, which have been found to play an essential role in the size and mass growth of massive SFGs at lower redshifts \citep{Rodriguez-Gomez2016, Mosleh2020, Pillepich2018, Kaw2021}.
	\begin{figure*}[ht!]
		\centering
		\includegraphics[width=0.9\textwidth]{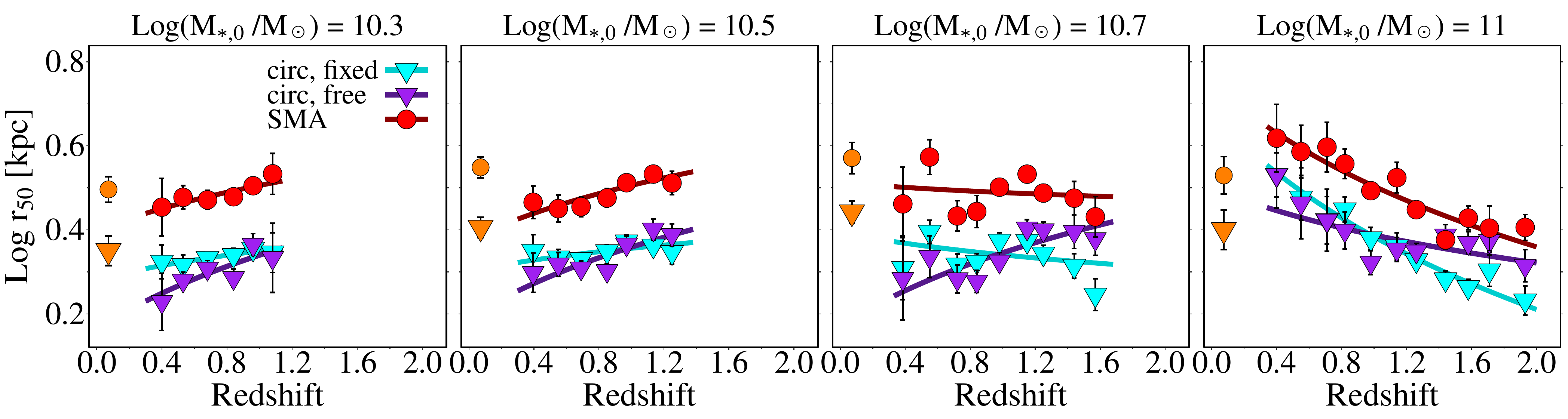}
		\caption{Comparing the redshift evolution of half-mass radius along the semi-major axis $r_{SMA}$ (red circles) with circularized half-mass size estimated from averaged ${b/a}$ of galaxy $r_{circ, fixed}$ (cyan down-pointing triangles) and circularized mass profiles fitting with free ${b/a}$ and PA at each radius $r_{circ, free}$ (purple down-pointing triangles) derived by \cite{Mosleh2020}. Although $r_{SMA}$ is larger than $r_{circ}$ and the median best-fit slopes are slightly different in some mass, there is no significant evolution in the half-mass sizes.}
		\label{fig11}
	\end{figure*}
	\subsection{Robustness of the size estimation} \label{subsec:robustness}
	Obtaining the half-mass sizes depends on the methods for converting the images to mass distribution, and then measuring the structural parameters from these maps. These processes involve many assumptions and also diverse methods of size measurement and, consequently cause uncertainties. Therefore, we briefly discuss the impacts of these factors in this section. 
	
	The complexities involved in estimating the distribution of stellar mass within galaxies make it challenging to calculate the stellar mass-weighted parameters. Due to these difficulties, various methods have been employed to derive the mass profiles. In previous studies, the estimation of stellar mass information has relied on either SED fitting \citep{Mosleh2017, Suess2019a, Mosleh2020} or color-M/L relations \citep{Suess2019a, Miller2023, vanderWel2024}. In addition, the mass profiles can be estimated directly from the 1D light profiles \citep{Mosleh2017, Suess2019a, Miller2023, vanderWel2024} or by creating 2D mass maps and finding the best-fit 1D profiles \citep{Suess2019a, Mosleh2020}. Employing diverse approaches, in combination with the uncertainties in SED fitting, or measuring the color-$M/L$ relationships, has the potential to introduce variation in the estimated mass-based sizes. Despite a few reported discrepancies that need more careful investigation \citep[e.g.][]{vanderWel2024}, it is interesting that there is an overall consistency among most studies regarding the evolution of half-mass size \citep{Mosleh2017, Suess2019a, Suess2019b, Mosleh2020, Hasheminia2022, Miller2023}. For instance, our results are robust by using mass-based sized from \cite{Suess2019a}. Therefore, it can be concluded that the discrepancy in the evolution of half-mass sizes compared to the light-based sizes cannot be attributed to uncertainties in observations.
	
	In addition, circularized sizes are frequently used in the literature. To ensure that the results are independent of size definition and methodology, we compared the evolution of the circularized half-mass radii with growth from semi-major axis sizes. In our analysis, two methods can be used to circularize the radius: one way is to fit fixed-center ellipses to the galaxy's isophotes or isomasses, while position angles ($PA$) and ellipticities ($e$) are free parameters. The profiles can be circularized at each radius using the corresponding axis ratio ($b/a$). Thus, the circularized mass- or light-weighted size is determined locally by fitting ellipses with free parameters ($r_{circ,free}$).
	
	The size of the galaxy can also be circularized based on its average axis ratio $(b/a)_{avg}$: 
	$r_{circ,fixed} = r_{SMA} \sqrt{(b/a)_{avg}}$
	We define the $(b/a)_{avg}$ as the average of axis ratio profiles in the range of $2r_{e,SMA}$ to the radius at which the stellar mass surface density reaches $10^7 \msun \mathrm{kpc}^{-2}$. Due to the differences in the axis ratio of inner and outer regions, the circularized half-mass radius obtained from these two methods shows some differences. As shown in Figure~\ref{fig11}, even though the $r_{e,SMA}$ is relatively larger, the slope of its redshift evolution is roughly in line with that of the circularized radii, particularly $r_{circ,fixed}$. Hence, the general evolution of the mass-based sizes is not affected by the way the sizes are estimated.
	
	Finally, given the discrepancy in the evolutionary paths of observational light and mass-based radii, more detailed and comprehensive studies are required to measure the size growth rates of galaxies and explore the underlying factors contributing to these differences in the literature. The high spatial resolution and NIR coverage of the James Webb Space Telescope (JWST), will allow us to measure mass-based structural parameters more precisely at higher redshifts and lower mass regimes. Additionally, it is crucial to undertake more precise analyses to identify the parameters that influence the more rapid evolution of the half-mass radii in the simulations compared to those estimated from observations. A simultaneous investigation of the color gradients and SFR profiles with advanced observational tools like the JWST can provide valuable insights into the complex mechanisms underlying the structural evolution of galaxies. Furthermore, combining these observations with state-of-the-art simulations will help to understand the existing discrepancies in the size evolution of simulated and observed galaxies.
	\section{Summary} \label{sec:summary}
	By analyzing the structural evolution and star formation activity of SFGs with a present-day stellar mass of $\log(M_{\ast,0}/\msun) = 10.3-11.0$, this study sheds light on the radial accumulation of stellar mass and the mechanisms that govern the assembly of different components since cosmic noon. In order to reduce biases resulting from employed methods for sample selection and parameter measurement, we attempted to track the structural evolution of each individual galaxy by utilizing parameters derived from the stellar mass maps. Our approach involves calculating the SMGH of SFGs with a given mass, followed by the identification of potential progenitors through the MSI method to reconstruct its evolutionary path. In addition, we have utilized the average SMGHs from the TNG50 simulations to corroborate our findings and directly compare them with simulations. The results indicate that there is no significant difference between the SMGHs calculated by these methods within the mass and redshift range of this study. Our key findings are summarized as follows:
	\begin{enumerate}
		\item In the case of intermediate to high-mass progenitors, the evolution of mass-weighted sizes exhibits shallower slopes compared to light-weighted sizes, underscoring the crucial role of color gradients in the analysis of galaxy evolution.
		\item Our results reveal that the evolution of SFGs depends on their final stellar mass. For intermediate and lower-mass SFGs ($\log(M_{\ast,0}/\msun) < 10.8$), the sizes of $r_{50}$ and $r_{80}$ remain almost unchanged over the past 10 Gyr, whereas $r_{20}$ tends to decrease, implying an increase in the concentration of the stellar mass distribution. On the other hand, at the massive end ($\log(M_{\ast,0}/\msun) > 10.8$), these two radii increase with cosmic time, leading to the $r_{50}$ and $r_{80}$ size evolution rates of $(1+z)^{-0.46\pm0.12}$ and $(1+z)^{-0.99\pm0.21}$. 
		\item The high-mass sample exhibits an inside-out growth pattern, whereas the intermediate and lower-mass SFGs accumulate their mass in a relatively self-similar manner.
		\item Even with similar SMGHs, the observed rates of size evolution differ from the predictions made by TNG50 simulations. The variations in feedback mechanisms, radial $M/L$ gradient, or mass redistribution may explain these differences.
		\item Our calculations, based on $r_e$ and halo mass from TNG50, reveal that the angular momentum remains relatively unchanged in simulations while the observed angular momentum experiences a significant reduction since $z\sim2$, which may result from the accretion of high-angular momentum gas in the simulations.
	\end{enumerate}
	Overall, our results highlight the complexity of galaxy structural evolution and the necessity of accurately measuring size growth rates to understand the differences between the observed light-based and mass-based radii, as well as the faster size evolution in simulations.
	\begin{acknowledgments}
		We thank the anonymous referee for the thorough review and valuable comments, which improved this manuscript. 
	\end{acknowledgments}
	\vspace{5mm}
	\facilities{HST, Sloan}
	\software{GALFIT \citep{Peng2010}, ggplot2 \citep{Wickham2016}, IRAF \citep{Tody1986, Tody1993}, iSEDfit \citep{Moustakas2013}, R \citep{R2021}, tidyverse \citep{Wickham2019}}
	\bibliographystyle{aasjournal}
	
\end{document}